\documentstyle[emulateapj]{article}

\newcommand{\teff}{\ifmmode{ T_{\rm{eff}}} \else $T_{\rm{eff}}$\fi} 

\newcommand{\halpha}{H$\alpha$}
\newcommand{\hbeta}{H$\beta$}
\newcommand{\hgamma}{H$\gamma$}

\newcommand{\heta}{H$\eta$}
\newcommand{\twid}{$\sim$}
\newcommand{\mh}{\ifmmode{ M_{\rm{H}} } \else $M_{\rm{H}}$\fi}
\newcommand{\mhe}{\ifmmode{ M_{\rm{He}} } \else $M_{\rm{He}}$\fi}

\begin{document}

\title{A Comparative Study of the Mass Distribution of \\
Extreme Ultraviolet-Selected White Dwarfs
\footnote{Spectral observations reported here were obtained with the
Multiple Mirror Telescope, a joint facility of the University of
Arizona and the Smithsonian Institution, and with the Bok telescope 
at the Steward Observatory of the University of Arizona.} 
}

\author{R.~Napiwotzki}
\affil{Dr.~Remeis-Sternwarte, Sternwartstr.~7, 96049 Bamberg, Germany}
\affil{email: {\em ai23@sternwarte.uni-erlangen.de}}

\author{Paul J. Green}
\affil{Harvard-Smithsonian Center for Astrophysics, 60 Garden St.,
Cambridge, MA 02138} 
\affil{email: {\em pgreen@cfa.harvard.edu}}
\and 

\author{Rex A. Saffer }
\affil{Dept. of Astronomy \& Astrophysics, Villanova University, 800
Lancaster Ave., Villanova, PA 19085}
\affil{email: \em{saffer@ast.vill.edu}}

\received{June 16, 1998}
\accepted{December 28, 1998}

\begin{abstract}
We present new determinations of effective temperature, surface 
gravity, and masses for a sample of 46 hot DA white dwarfs
selected from the EUVE and ROSAT Wide Field Camera bright
source lists in the course of a near-IR survey for low mass
companions.  Our analysis, based on hydrogen NLTE model
atmospheres, provides a map of LTE correction vectors, which allow
a thorough comparison with previous LTE studies.  We find 
previous studies underestimate both the systematic errors and the
observational scatter in the determination of white dwarf
parameters obtained via fits to model atmospheres.

The structure of very hot or low mass white dwarfs depends 
sensitively on their history. To compute white dwarf masses,
we thus use theoretical mass-radius relations 
that take into account the complete evolution from the main sequence. 
We find a peak mass of our white dwarf sample of $0.59M_\odot$,
in agreement with the results of previous analyses. 
However, we do not confirm a trend of peak mass with temperature
reported in two previous analyses.  

Analogous to other EUV selected samples, we note a lack of low mass
white dwarfs, and a large fraction of massive white dwarfs.  
Only one white dwarf is likely to have a helium core. 
While the lack of helium white dwarfs in our sample can be easily 
understood from their high cooling rate and therefore low detection
probability in our temperature range, this is not enough to explain the
large fraction of massive white dwarfs.  This feature very likely
results from a decreased relative sample volume for low mass white
dwarfs caused by interstellar absorption in EUV selected samples. 

\end{abstract}


\section{Introduction}

UV observations of EUV-detected stars have revealed
the presence of about 15 hot white dwarf companions to bright stars in
non-interacting binary systems  (e.g., Burleigh, Barstow, \& Fleming
1997).  At optical wavelengths, these WDs are hidden because of their
close proximity to much more luminous companions which are main
sequence (spectral type K or earlier) or evolved stars. 

A fascinating variety of objects are known or proposed to contain
white dwarf (WD) stars in interacting binary systems.  A partial list
includes novae, cataclysmic variables, symbiotic stars, Ba and CH
giants, Feige~24-type systems and dwarf carbon stars (Green \& Margon
1994). These systems offer great insights to evolution and dynamical 
processes in binaries. 

A number of interacting binary systems where the white dwarf is the primary
(i.e., optically brightest) star have also been found among
EUV-detected systems (e.g., 6 close, interacting white dwarf/red dwarf
binaries by Vennes \& Thorstensen 1994).  Optical or ultraviolet
spectral observations are most commonly used to detect companions to
WD primaries, by searching for (1) the presence of narrow Balmer line
emission overlying the broad smooth Balmer absorption of the WD, (2) a
composite white dwarf + main sequence spectrum, or (3) radial velocity (RV)
variations.  However, only WDs with very close, or intrinsically
active companions will be found by method (1).  For hot white dwarf systems,
composite spectra (2) are only expected to be visible if the
companion's spectral type is early enough. RV variations (3) require
multiple observations at high spectral resolution, and detection
strongly favors close and/or massive companions.

All the discoveries mentioned above have been strongly dominated by
these selection effects, with companions biased to earlier types than
predicted by the simulations of deKool \& Ritter (1993) and others.
Scaling from deKool \& Ritter's results, Vennes \& Thorstensen (1994)
estimate that ``at least twice as many close binary systems remain to
be identified from EUV surveys, most of them with a low mass
secondary.'' The resulting sample of binaries known to date therefore
must diverge strongly from the intrinsic distribution, in overall
normalization, as well as in mass and spectral type of the main
sequence companions.

The current study, conceived as a complement to optical studies, 
began as a near-IR photometric survey for low mass
companions to hot white dwarfs (WDs).  By investigating only
EUV-detected WDs, we obtain a very reasonably-sized but complete
sample of young WDs, next to which very late-type dwarf  
companions can be detected in the near-infrared by searching
for a $K$ excess over that expected from the white dwarf.
Many hot white dwarfs ($T_{\rm eff}>24000$K; Finley et al. 1993) have
been detected in the recent EUV all-sky surveys.  EUV detection of
these hot WDs depends primarily on their temperature, distance, and
the intervening Galactic ISM.  Our sample of EUV WDs (whose selection
we define below) offers excellent flux contrast in the IR relative to
optical; cool companions will almost always be brighter in the $K$
band than the hot WDs.  

To know what $K$ mag to expect for the WDs, we benefit from 
constraints on log\,$g$, radius, and $T_{\rm eff}$ derivable
from optical spectra for the WDs in our sample, using 
NLTE model atmosphere fits (Napiwotzki et al.\ 1993, Napiwotzki 1997).
The resulting predictions for $K$ magnitudes allow a direct search for any
IR excess from a cool companion.  In some cases, IR colors will also
provide a preliminary spectral type.  Results from the IR survey 
will be presented in an upcoming paper.

An additional motivation is the study of the white dwarf mass
distribution. Since the pioneering work of Koester, Schulz \&
Weidemann (1979; KSW) it is well established that the masses of white
dwarfs cluster in a narrow range around $0.6M_\odot$, remarkable
given that white dwarfs stem from progenitors with masses
ranging from below $1M_\odot$ up to $\approx $$8M_\odot$. Precise
knowledge of the white dwarf mass distribution puts constraints on the
theory of stellar evolution, especially the poorly understood mass
loss process during the final stages of stellar evolution. With two
recent exceptions (Beauchamp et al.\ 1996 and Dreizler \& Werner 1996, 
who analyzed samples of
helium-rich DB and DO white dwarfs, respectively) the mass distribution  
has only been determined
for hydrogen-rich DA white dwarfs. However, this is not a severe
limitation, because this spectral class comprises about 80\,\% of all
known white dwarfs.

The analysis of KSW, along with other follow-up investigations in the
early eighties, used photometric data of which the Str\"omgren and
Greenstein multichannel colors were the most suitable. Both systems
provide temperature- and gravity-sensitive indices. Alternatively, KSW
and others used trigonometric parallax measurements to directly
calculate the stellar radius. However, the latter method is practical
only for a small sample and suffers from considerable measurement
uncertainty. Unfortunately, the photometric indices have their highest
sensitivity near 10000\,K. At this temperature DA white dwarfs have a
convective atmosphere and the results depend critically on the adopted
parameters of mixing length theory.

The situation improved at the beginning of the nineties, when the
development of modern, highly efficient detectors made it possible to
obtain high-quality spectra of large numbers of white dwarfs and
determine the stellar parameters from a fit to the detailed profiles
of the Balmer lines. This method yields sufficient accuracy for white
dwarfs hot enough to have a radiative envelope. The first
comprehensive sample of white dwarfs analyzed by this method was
presented by Bergeron, Saffer \& Liebert (1992; hereafter BSL). 
As attributed to higher precision of spectroscopic methods,
this investigation yielded a white dwarf mass distribution even
narrower than found by KSW and other previous studies. 

At the same time, Kidder (1991) analyzed a sample of hot DA white
dwarfs discovered through positional coincidences of catalogued hot DA
white dwarfs in existing soft X-ray databases. Three soft x-ray
sources corresponding to white dwarfs were found, having relatively
low effective temperatures, $\approx$25,000\,K, determined independently
using complementary optical and UV spectroscopy. Kidder et al. (1992)
analyzed an expanded sample to derive photospheric He abundances for
the hotter objects and establish an effective observational
low-temperature threshold for the detection of pure hydrogen DA white
dwarfs at soft X-ray wavelengths.

In 1997 three groups (Marsh et al.\ 1997, M97; Vennes et al.\ 1997,
V97; Finley, Koester \& Basri 1997, FKB) published results on the mass
distribution of Extreme Ultraviolet (EUV) selected white dwarfs. Due
to the selection criterion, these samples contain the hottest white
dwarfs (\teff$ > 25000$\,K), as cooler white dwarfs do not emit
significant EUV radiation. The derived mass distributions in the
EUV-selected samples are similar to that of BSL, but show some
interesting deviations in detail. The frequency of very high mass
white dwarfs is much larger, and that of very low mass white dwarfs
much smaller, than 
in BSL. These findings can at least partly be explained by selection
effects (see the discussion in FKB). More serious is a trend of the
peak mass with temperature. 
V97 found that their mass distribution peaks at $0.598M_\odot$, while 
the BSL distribution peaks at $0.568M_\odot$, with masses 
computed using Wood's (1995) mass-radius 
relation with ``thick'' layers ($\mh = 10^{-4}M_{\rm{WD}}$, 
$\mhe = 10^{-2}M_{\rm{WD}}$).   This discrepancy diminishes
slightly, if the ``very thin layer'' ($\mhe = 10^{-4}M_{\rm{WD}}$,
no hydrogen layer) 
mass-radius relations are used (peak masses of $0.556M_\odot$ and 
$0.532M_\odot$ for the V97 and BSL sample, respectively).
V97 interpreted this as
evidence for a very thin hydrogen layer of the DA white
dwarfs. However, the effects are small so this result depends
strongly on the accuracy of the derived stellar parameters.

FKB estimated the internal accuracy of different analysis methods from
Monte Carlo simulations. The precision reachable by Balmer line
fitting is very compelling: $\Delta$\teff/\teff$< 0.01$ for \teff$ <
60000$\,K. However, for spectra with very high signal-to-noise ratios
(S/N), errors introduced by details of the observation and
reduction techniques 
(e.g., extraction, flat fielding, flux and wavelength calibration)
might be more important, but are very difficult to determine. 
Additionally, one has to take into account differences in the model
atmosphere calculations and fitting procedure. Together with the 
results presented in this paper, we now have four samples of hot white
dwarfs, analyzed in a similar way, and with significant overlap. This
offers the opportunity to determine the real accuracy of the 
spectral analysis of hot white dwarfs, including many possible
systematic effects.  

We present the selection criteria of our sample in Sect.~2 and the
observations and data reduction procedures in Sect.~3. Details on our
model atmospheres are given in Sect.~4. The results and a detailed
comparison with the previous analyses of EUV selected white dwarfs are
presented in Sect.~5.  We finish with a discussion of our results and
an outlook.

\section{Sample Selection Criteria}

We chose to limit our uniform sample to DAs, for which models provide the
best temperature and mass constraints. We start with 73 known DA white
dwarfs in the EUVE bright source list (Malina et al. 1994).  Excluding
sources at low galactic latitudes ($|b|<15$) and southerly
declinations ($\delta<-20$) yields a list of 28 DAs.  A similar
procedure for non-overlapping DAs listed in the {\it ROSAT} Wide Field
Camera survey Bright Source Catalogue (Pounds et al. 1993) yields 29
objects.

We have removed from our uniform sample 2 well-known stars with 
published sensitive optical spectrophotometry and IR photometry
(Feige~24 and HZ~43).   Two DAs with broad line profiles due to
magnetic splitting were also excluded (PG\,1658+441
and PG\,0136+251).  Eight known binaries are also
excluded from the uniform sample:
V471~Tau
	(Vennes, Christian, \& Thorstensen 1998), 
PG\,0824$+$289, 
	(Heber et al. 1993)
HD\,74389B
	(Liebert, Bergeron, \& Saffer 1990), 
RE\,J1016$-$052
	(V97), 
PG\,1033+464 (GD\,123),
	(Green, Schmidt, \& Liebert 1986),
RE\,J1426$+$500
	(V97),  
RE\,J1629+780
	(Catalan et al. 1995), and 
IK~Peg
	(Wonnacott et al. 1993), 
leaving 47 objects.  

In this paper, a handful of objects which fell outside the 
uniform sample definitions just outlined were included for
observation.  These include the known binaries PG\,1033+464 and
RE\,J1629+780 and the magnetic white dwarf PG\,1658+441, as well as
MCT\,0455-2812 which was outside the sample declination limits.

Due to observing constraints (a combination of weather, poor seeing
and faint objects, or celestial placement of objects) no spectra 
were obtained for sample objects RE\,J0443-034, RE\,J0916-194,
or PG1040+451.  PG1234+482 was originally classified as a sdB star
and thus excluded; it has since been reclassified as a DA (Jordan et
al. 1991).  The final sample we analyze here thus includes 46 DA white
dwarf stars for which we present new model NLTE fits to optical
spectra. 

We note that since the sample selection was performed, several relevant
discoveries pertaining to sample objects have been made.
RE\,J0134-160 (GD984) has central Balmer emission components produced by a
dMe companion.  RE\,J1440+750 turns out to be a magnetic DA (Dreizler et
al. 1994).  The uniform sample will be discussed in a followup
paper treating the IR photometry and binary fraction.

\section{Observations}

Dates are listed for all observations in Table~\ref{t:results}.
On the nights of 04 - 06 January 1996 we obtained spectra at Steward
Observatory's Kitt Peak Station using the Bok 2.3-m reflector equipped
with the Boller \& Chivens 
Cassegrain spectrograph and UV-flooded Loral 800$\times\/$1200 CCD.
Most spectra were dispersed with a 600 $l$/mm first-order grating used
behind a 4.5$^{\prime\prime}\times~$4$^\prime$ long slit. The
instrumentation provided wavelength coverage 
$\lambda\lambda$3400--5600 at a spectral resolution of \twid 5\AA\
FWHM. On the last night of the observing run we employed a new 400
l/mm grating providing wavelength coverage $\lambda\lambda$3500--6790
at a spectral resolution of \twid 7\AA\ FWHM.  

We also obtained spectra at the MMT on Mt Hopkins, 09 - 11 May 1996, using
the 300~$l$/mm grating in first order on the Loral 3k$\times$1k CCD of
the Blue Channel spectrograph.  This yields coverage from about 3350
to 8800~\AA, and the 2$^{\prime\prime}$ slit width we used resulted in a
spectral resolution of about 4~\AA\, FWHM.   Several objects
were kindly obtained for us using the identical instrumental
configuration at the MMT by Perry Berlind on 08 April 1997.

Exposures at both telescopes ranged from one to thirty minutes for
program stars, and for all observations the long slit was rotated to
the parallactic angle according to the calculations of Filippenko
(1982). The airmasses were held below 1.5 in almost all cases.
All spectra were extracted from the two-dimensional images and reduced
to linear wavelength and intensity scales using standard reduction
packages in the Image Reduction and Analysis Facility (IRAF). These
operations included bias subtraction, flat field division by images
obtained by exposing on dome or internal quartz lamps, centroiding and
summation of the stellar traces on the two-dimensional images, sky
subtraction, wavelength calibration using spectra of He/Ar arc lamps,
and absolute flux calibration using the flux standards of Massey et
al. (1988). Details of these reduction procedures are given
by BSL.

We present in Figs.~\ref{f:steward} and~\ref{f:MMT} our collection of white 
dwarf spectra. The spectra obtained at the Steward observatory 2.3\,m
telescope  are found in Fig.~\ref{f:steward}, and the MMT spectra in
Fig.~\ref{f:MMT}.  Some stars observed repeatedly appear in both figures.
The signal-to-noise ratio of our spectra ranges from 35 to 200 with an
average of 90. Ninety percent of our spectra have a S/N of at least 45. 

\section{Model atmospheres}

We calculate hydrogen model atmospheres with the NLTE code
developed by Werner (1986). Basic assumptions are those of static,
plane-parallel atmospheres in hydrostatic and radiative equilibrium.
In contrast to the atmospheres commonly used to analyze DA white
dwarfs, we relax the assumption of local thermal equilibrium (LTE) and
solve the detailed statistical equilibrium instead. As described in
Werner (1986), the accelerated lambda iteration (ALI) method is used
to solve the set of non-linear equations. The impact of NLTE on
white dwarf atmospheres is discussed in detail in Napiwotzki (1997).

\ion{H}{1} levels and lines are included in NLTE up to $n=16$. Line
blanketing by the Stark broadened hydrogen lines is taken into account
consistently.  As the hydrogen atmospheres of DA white dwarfs are
stable for \teff$ > 15000$\,K, convection is not included in our
atmospheric models. Pressure dissolution of the higher levels is
described by  the Hummer \& Mihalas (1988) occupation probability formalism
following the NLTE implementation by Hubeny et al.\ (1994). The
synthetic spectra are computed with the extended VCS broadening tables
(Vidal et al.\ 1970) provided by Sch\"oning \& Butler (priv.\ comm.)
and Lemke (1997). We followed the prescription of Bergeron (1993) and
increased the critical ionizing field adopted to calculate the occupation
probability by a factor two. The motivation is not a flaw in the
Hummer \& Mihalas (1988) formalism, but a compensation for the 
inadequacy of the standard Stark broadening theory when line wings overlap.
Our NLTE model grid covers the temperature range
17000\,K $<$ \teff$ < $100000\,K (stepsize increasing with \teff\, from
2000\,K to 10000\,K) and gravity range $6.50 < \log g < 9.75$
(stepsize 0.25).

Although deviations from LTE are small for most DA white dwarfs they
become significant for the hottest stars in our sample (cf. Napiwotzki
1997). Since we intend to compare our results with three other samples
analyzed by means of LTE atmospheres, we have produced a map with LTE
correction vectors.  For this purpose we calculated a set of LTE model
atmospheres using the technique described in Napiwotzki (1997) of
drastically enhancing collisional rates between the atomic levels in
the NLTE code. This forces the occupation numbers to be in LTE and
guarantees consistency with the NLTE atmospheres.

The synthetic LTE spectra were transformed into ``observed'' spectra by
convolving them with a Gaussian of 5\,\AA\ FWHM, rebinning them to 2\,\AA\,
and adding Poisson noise corresponding to a continuum S/N of 100. These
simulated spectra were analyzed with the NLTE grid following the 
procedure outlined in Sect.~\ref{s:analysis}. 1000 simulations were run for
every parameter set to eliminate the effect of random errors.
The resulting offsets are displayed in
Fig.~\ref{f:ltevectors}. The orientation of the vectors corresponds to the
correction, which must be applied to transform LTE results to the NLTE scale.

As expected from the results of Napiwotzki (1997) the differences are
negligible for DA white dwarfs cooler than $\approx$40000\,K, but can
be significant for hotter stars. The LTE vectors show the
trend that NLTE effects increase with increasing temperature and
decreasing gravity (see e.g.\ Fig.~3 and~4 in Napiwotzki, 1997, for
the case of DA white dwarfs). The small corrections found for the
models with the highest temperature and lowest gravity seems to
contradict this behavior. A look at the line profiles reveals that the
NLTE deviations are larger for, say, the \teff$= 90000$\,K, $\log g =
7.0$ model than the 70000\,K, 8.0 model. However, the correction
vectors are not a simple function of NLTE deviations measured e.g.\ as
equivalent width difference, but depend also on the way line profiles
vary with temperature and gravity. The small corrections found for the
high temperature/low gravity models are produced by the cancellation
of these effects. We expect a reliable transformation between the LTE
and NLTE temperature scales only if the deviations are not too
large. A conservative upper limit is $\approx$70000\,K.  A thorough
comparison with the previous LTE analyses is presented in
Sect.~\ref{s:comparison}.

We checked our models by a comparison of our LTE spectra with some DA 
model spectra kindly provided by D.~Koester. Model parameters were
$\teff = 30000$\,K to 70000\,K in 10000\,K steps and $\log g = 8.0$ for 
pure hydrogen models. Input physics are very similar. 
In particular both model calculations adopt twice the critical ionizing field
for the calculation of line profiles. We treated Koester's model spectra
the same way we treated the LTE spectra above and fitted them with our
LTE grid. The result was quite satisfactory: the temperature
differences were always below 1.5\% and the gravity differences never
exceeded 0.03\,dex.  

Metals were ignored in our calculations, but they can modify the hydrogen line
profiles by their effect on the atmospheric structure. Lanz et al.\ (1996)
analyzed the Balmer lines of the hot DA G\,191\,B2B with pure hydrogen LTE and
NLTE atmospheres and an NLTE model with full metal line blanketing. They
concluded that the effect of metal line blanketing on the Balmer lines was
relatively small, and the difference between LTE and NLTE was found to be the
most important effect. A recent study by Barstow et al.\ (1998), which
investigated several hot white dwarfs in the temperature range around 
60000\,K derived larger metal line blanketing effects of the order of the NLTE
effects. Since the LTE analyses of M97, V97, and FKB are based on pure hydrogen
models, our results should be consistent with theirs in any case. 

\section{Spectral analysis and results}
\label{s:analysis}

Atmospheric parameters of our DA white dwarfs are obtained by simultaneously
fitting line profiles of the observed Balmer lines with the NLTE model spectra
described above. We use the least-square algorithm described in BSL. The
observed and theoretical Balmer line profiles are normalized to a linear
continuum (both spectra are $F_\lambda$) in a consistent manner.  Wavelength
shifts are determined with a cross-correlation method and applied consistently
to each complete spectrum.  The synthetic spectra are convolved to the
observational resolution with a Gaussian and interpolated to the actual
parameters with bicubic splines, and interpolated to the observed wavelength
scale. 

The atmospheric parameters \teff\, and $\log g$ are then determined by
minimizing the $\chi^2$ value by means of a Levenberg-Marquardt steepest
descent algorithm (Press et al.\ 1986). Several tests revealed that our
interpolation routine is rather robust concerning spacing of our model 
grid and yields reliable results even at the edge of the model grid.

Finally, an estimate of the internal errors can be derived from the
covariance matrix. In contrast to BSL, we estimate the noise of the
spectra ($\sigma$) used for the $\chi^2$ fit from the neighboring
continuum of each line. The S/N is adopted to be constant throughout the
line.

\begin{deluxetable}{rlr@{$\pm$}lr@{$\pm$}lrl}
\tablewidth{0cm}
\tablecaption{ATMOSPHERIC PARAMETERS OF PROGRAM STARS
\label{t:results}}
\tablehead{
 \colhead{RE\,J}
&\colhead{Other Names}
&\multicolumn{2}{c}{\teff}
&\multicolumn{2}{c}{$\log g$}
&\colhead{$M/M_\odot$}
&\colhead{Observation} \\
\colhead{ } 
&\colhead{ }  
&\multicolumn{2}{c}{ }
&\multicolumn{2}{c}{ }
&\colhead{ } 
&\colhead{Date}
}
\startdata
0007+331	&GD\,2		&46493&514	&7.83&0.05	&0.602	& 6 Jan 1996 \\
0134$-$160\tablenotemark{a}&GD\,984, PHL\,1043&44866&667 &7.77&0.05	&0.572	& 5 Jan 1996 \\
0237$-$122	&PHL\,1400	&32077&177	&8.45&0.04	&0.890	& 5 Jan 1996 \\
0348$-$005	&GD\,50		&39508&464	&9.07&0.06	&1.215	& 4 Jan 1996 \\
0427+740	&		&48587&1044	&7.93&0.08	&0.646	& 5 Jan 1996 \\
0457$-$280	&MCT\,0455$-$2812&51199&786	&7.72&0.05	&0.570	& 4 Jan 1996 \\
0512$-$004	&		&31733&139	&7.40&0.03	&0.458	& 5 Jan 1996 \\
0521$-$102	&		&33186&301	&8.60&0.06	&0.980	& 5 Jan 1996 \\
0841+032	&		&38293&252	&7.75&0.03	&0.544	& 4 Jan 1996 \\
0902$-$040	&		&23218&160	&7.84&0.02	&0.544	& 4 Jan 1996 \\
0907+505	&PG\,0904+511	&32167&338	&8.11&0.07	&0.695	& 4 Jan 1996 \\
0940+502	&PG\,0937+506	&36034&283	&7.69&0.04	&0.519	& 9 Apr 1997 \\
0957+852	&		&51311&1348	&8.37&0.10	&0.866	& 5 Jan 1996 \\
1019$-$140	&		&31524&102	&7.92&0.02	&0.606	& 9 Apr 1997 \\
1029+450	&PG\,1026+454	&35518&247	&7.70&0.04	&0.520	& 9 Apr 1997 \\
1032+532	&		&43587&506	&7.95&0.05	&0.644	& 4 Jan 1996 \\
1033$-$114	&G\,162$-$66, LTT\,3870	&24685&252&7.85&0.03	&0.553	& 4 Jan 1996 \\
1036+460\tablenotemark{e}	&GD\,123	&29361&251	&8.02&0.05	&0.648	& 4 Jan 1996 \\
1043+490\tablenotemark{e}	&	&41132&1178	&7.94&0.13	&0.635	& 5 Jan 1996 \\
1044+574	&PG\,1041+580	&30338&153	&7.81&0.03	&0.550	& 4 Jan 1996 \\
1100+713	&PG\,1057+719	&41104&814	&7.84&0.09	&0.593	& 4 Jan 1996 \\
1112+240	&Ton\,61	&39824&636	&7.78&0.07	&0.563	& 5 Jan 1996 \\
1122+434	&PG\,1120+439	&26996&151	&8.31&0.02	&0.803	& 9 Apr 1997 \\
1126+183\tablenotemark{e}&PG\,1123+189	&54334&1983	&7.76&0.13	&0.594	& 4 Jan 1996 \\
1128$-$025	&PG\,1125$-$026	&30699&380	&8.24&0.08	&0.767	& 4 Jan 1996 \\
1148+183	&PG\,1145+188	&25758&299	&7.91&0.04	&0.587	& 4 Jan 1996 \\
1235+233\tablenotemark{b}&PG\,1232+238 &46569&523&7.83&0.05	&0.602	& 11+12 May 1996 \\
1257+220\tablenotemark{b}&GD\,153 &38926&142	&7.78&0.02	&0.560	& 4 Jan 1996, 12 May 1996 \\
1336+694	&PG\,1335+701	&29607&87	&8.34&0.02	&0.824	& 10 May 1996 \\
1431+370	&GD\,336	&34404&115	&7.91&0.02	&0.608	& 11 May 1996 \\
1446+632	&		&37947&254	&7.79&0.04	&0.562	& 11 May 1996 \\
1629+780\tablenotemark{c}&	&41043&338	&7.92&0.04	&0.627	& 11 May 1996 \\
1638+350	&PG\,1636+351	&35404&142	&7.98&0.02	&0.642	& 10 May 1996 \\
1643+411\tablenotemark{b}&PG\,1642+414 &28815&81&8.22&0.02	&0.753	& 9 May 1996 \\
1650+403\tablenotemark{b}&	&38144&211	&7.97&0.03	&0.643	& 9 May 1996 \\
1711+664\tablenotemark{d}&	&48989&757&8.89&0.06	&1.141	& 12 May 1996, 9 Apr 1997\\
1726+583\tablenotemark{b}&PG\,1725+586 &53561&542&8.23&0.04	&0.795	& 9 May 1996 \\
1800+683	&KUV\,18004+6836&44723&424	&7.80&0.04	&0.585	& 10 May 1996 \\
1820+580	&		&44099&264	&7.78&0.03	&0.574	& 9 May 1996 \\	
1845+682	&KUV\,18453+6819&36120&189	&8.23&0.03	&0.770	& 10 May 1996 \\
2116+735\tablenotemark{b}&KUV\,21168+7338 &50812&354&7.72&0.03	&0.569	& 5 Jan 1996, 9 May 1996 \\
2207+252	&		&26964&174	&8.27&0.03	&0.779	& 5 Jan 1996 \\
2312+104	&GD\,246	&53088&968	&7.85&0.07	&0.624	& 4 Jan 1996 \\
\enddata
\tablenotetext{}{NOTES: The error estimates were taken from the
$\chi^2$ procedure.}
\tablenotetext{a}{spectrum slightly contaminated by cool companion; 
		the cores of \hbeta\ and \hgamma\ are excluded from the fit}
\tablenotetext{b}{Weighted mean of individual observations}
\tablenotetext{c}{red part of spectrum contaminated by cool companion; 
		\hbeta\ and the core of \hgamma\ excluded from fit}
\tablenotetext{d}{Close-by optical companion; spectrum apparently not 
		contaminated}
\tablenotetext{e}{red part of spectrum contaminated by cool companion; 
		\hbeta\ excluded from fit}
\end{deluxetable}

\clearpage

The results are given in Table~\ref{t:results}, with illustrative examples
shown in Fig.~\ref{f:sampleres}. We adopt, for the moment, the usual
practice and indicate in 
Table~\ref{t:results} the  internal errors estimated from the quality
of the $\chi^2$ fit. However, one should keep in mind that these 
errors can only serve as lower limits.  We will show below
(Sect.~\ref{s:comparison}) that these formal errors derived from the
$\chi^2$ fit significantly underestimate the real errors. 

External errors can be estimated from multiple observations and
analysis of the same star.  We obtained repeat observations for
a subsample of 6 stars, for which results are given in
Table~\ref{t:repeat}. The gravity values of all six stars agree within
the estimated internal errors. The same is true for four temperature
comparisons, too. However, the differences found for RE\,J1650+403 and
RE\,J2116+735 are significantly larger. This is in line with the
external errors we estimate from a comparison with the studies of M97,
V97, and FKB (see Sect.~\ref{s:comparison}). 

\begin{deluxetable}{rr@{$\pm$}lr@{$\pm$}lr@{$\pm$}lr@{$\pm$}l}
\tablewidth{0cm}
\tablecaption{RESULTS of REPEATED OBSERVATIONS 
\label{t:repeat}}
\tablehead{
&\multicolumn{4}{c}{Observation 1}
&\multicolumn{4}{c}{Observation 2}
\\
\colhead{RE}
&\multicolumn{2}{c}{\teff}
&\multicolumn{2}{c}{$\log g$}
&\multicolumn{2}{c}{\teff}
&\multicolumn{2}{c}{$\log g$}
}
\startdata
1235+233	&46308&837	&7.85&0.08 	&46737&670	&7.82&0.06\\
1257+220	&39349&317	&7.76&0.04	&38820&159	&7.78&0.02\\
1643+411	&28813&120	&8.23&0.02	&28818&111	&8.21&0.02\\
1650+403	&37798&245	&7.94&0.04	&39126&413	&8.01&0.05\\
1726+583	&52712&927	&8.27&0.06	&54003&669	&8.20&0.05\\
2116+735	&50131&384	&7.71&0.03	&54604&906	&7.76&0.06\\
\enddata
\tablenotetext{}{NOTES: The error estimates were taken from the
$\chi^2$ fitting procedure.}
\end{deluxetable}

\subsection{Binaries}

Five stars in our sample, RE\,J0134-160, RE\,J1036+460, RE\,J1043+490, 
RE\,J1126+183, and RE\,J1629+780, show clear
signs of binarity in the red part of their spectra. Two more stars, 
RE\,J1711+664 and RE\,J2207+502, are members of visual binaries.

\paragraph{RE\,J1629+780:}
The red part of the spectrum of RE\,J1629+780 is heavily contaminated
by a M type main sequence companion. The composite spectrum and the
spectrum of the M star after subtracting the white dwarf component are
shown in Fig.~\ref{f:compspek}. The characteristic bands of TiO are
easily recognizable. Catal\'an et al.\ (1995) determined a spectral
type dM4.  The \halpha\ and \hbeta\ lines are seen in emission, which
indicates a chromospherically active Me star. Sion et al.\
(1995) detected a flare-like increase of the Balmer line emission.

The comparison in Fig.~\ref{f:compspek} demonstrates that the blue
part ($\lambda < 4400$\,\AA) of the white dwarf spectrum is not
disturbed by the M dwarf. Thus, we excluded \hbeta\ and the core of
\hgamma\ from the fit and derived the parameters \teff$ = 41000$\,K
and $\log g = 7.92$ from \hgamma\ to \heta. These parameters are in
reasonable agreement with the results of Catal\'an et al.\ (\teff$ =
41800$\,K; $\log g = 8.0$), who however included \hbeta, and Kidder
(1991; \teff$ = 42500$\,K, $\log g = 7.6$), who fitted Lyman-$\alpha$
and the Balmer lines \hbeta\ and \hgamma.

\paragraph{RE\,J1036+460, RE\,J1043+490, RE\,J1126+183:} 
Three more white dwarfs of our observed sample are known binaries
(RE\,J1036+460, RE\,J1126+183: Green, Schmidt, \& Liebert 1986,
RE\,J1043+490: Schwartz et al.\ 1995) and show red excesses in our
spectra:  Since the \hbeta\ lines of the white dwarfs are contaminated
they were excluded from the fit. \hgamma\ and higher Balmer lines are
virtually uncontaminated. A discussion of these stars and newly
discovered binaries will be given in a forthcoming paper.  

\paragraph{RE\,J0134-160 (GD\,984):}
Although our spectrum ends at only 5600\,\AA\ a red excess is obvious.
Subtraction of the theoretical white dwarf flux leaves an M star
spectrum.  The M dwarf contribution is much smaller than in
RE\,J1629+780 and we excluded only the cores of \hbeta\ and \hgamma,
filled in by the Balmer line emission, from the fit. Bues \& Aslan (1995)
suspected a hot third component in RE\,J0134-160.  However,
outside of the \hbeta\ and \hgamma\ cores, the Balmer lines are well
reproduced by  our best fit without any indication of a third
component.  Indeed, a subdwarf component as suggested by Bues \& Aslan
(1995) is almost certainly not present, since its flux would dominate
in the blue.

\paragraph{RE\,J1711+664:} Since a late type star is separated from the white 
dwarf by only $2\farcs 5$, we took care to get an uncontaminated
white dwarf spectrum. We obtained a spectrum under good seeing conditions.
No excess is present up to the red limit at 8500\,\AA. 

\paragraph{RE\,J2207+252:}
This white dwarf has a red companion $8\farcs 5$ away. Schwartz et al.\ (1995)
estimated spectral type K4V and distance 65\,pc from its colors. With the
parameters from Table~\ref{t:results} a white dwarf distance of 62\,pc 
results. Thus it is likely that both stars form a physical pair. The angular
separation corresponds to $\approx$500\,AU.

\subsection{Magnetic white dwarfs} 

Two white dwarfs of our sample show Zeeman splitting of the Balmer
line cores, indicative of a magnetic field. RE\,J1659+440 (PG\,1658+441)
is a well known star, already analyzed by Schmidt et al.\ (1992).  The
discovery of the magnetic nature of RE\,J1440+750 (HS\,1440+7518) was
announced by Dreizler et al.\ (1994; note the naming confusion
corrected in Dreizler et al.\ 1995: 
HS\,1412+6115 should have been HS\,1440+7518). 
Although RE\,J1440+750 was analyzed by V97,
they did not remark on its magnetic nature. This
is likely due to the lack of coverage of the \halpha\ line, which
displays the most pronounced Zeeman effect.

Flux calibrated spectra of the magnetic white dwarfs are displayed in
Fig.~\ref{f:magwd}. The PG\,1658+441 analysis of Schmidt et al.\
(1992) resulted in \teff$ = 30500$\,K and $\log g = 9.35$. The
magnetic splitting was best reproduced by a 3.5\,megagauss (MG) dipole
inclined $60^\circ$ to the line of sight (producing a mean surface
field strength $B_{\rm{S}} = 2.3$\,MG). From the linear Zeeman
effect we estimated a mean magnetic strength of 8\,MG for RE\,J1440+750,
consistent with the estimate given in Dreizler et al.\ (1994).

The temperature and gravity of RE\,J1440+750 were derived from a fit of
the higher Balmer lines \hgamma\ to \heta, which are less affected by
the magnetic splitting. Results are given in Table~\ref{t:magwd}
supplemented by the Schmidt et al.\ (1992) fit of PG\,1658+441. Our
own fit gave similar results albeit with lower accuracy.  Our fits of
RE\,J1440+750 can only provide a rough estimate of the 
stellar parameters. Accurate results can only be expected from a
detailed treatment of the magnetic effects.

\begin{deluxetable}{rlr@{$\pm$}lr@{$\pm$}lrrl}
\tablewidth{0cm}
\tablecaption{PARAMETERS of the MAGNETIC WHITE DWARFS
\label{t:magwd}}
\tablehead{
\colhead{RE}
&\colhead{other names}
&\multicolumn{2}{c}{\teff}
&\multicolumn{2}{c}{$\log g$}
&\colhead{$M/M_\odot$}
&\colhead{$B_{\rm{S}}$}
&\colhead{Observation}\\
\colhead{ }
&\colhead{ }
&\multicolumn{2}{c}{ }
&\multicolumn{2}{c}{ }
&\colhead{ }
&\colhead{ }
&\colhead{Date}
}
\startdata
1440+750	&HS\,1440+7518	&36154&275	&8.87&0.05	&1.128	&7.7
		&10 May 1996\\
1659+440\tablenotemark{a}&PG\,1658+441 &30510&200&9.36&0.07	&1.311	&2.3
		&10 May 1996\\
\enddata

\tablenotetext{a}{Parameters from Schmidt et al.\ 1992}
\end{deluxetable}

\subsection{Internal, external, and systematic errors}
\label{s:comparison}

We have now presented the results of a homogeneous analysis of a
sample of 46 hot, EUV selected white dwarfs based on Balmer line
fitting. Three other large samples analyzed with the same method were
recently published by M97, V97, and FKB. Since considerable overlap
exists between all four samples, this allows a direct check for
systematic errors and the individual scatter on a star by star basis
for white dwarfs hotter than 25000\,K.

Since in contrast to previous works, our analysis is based on NLTE model
atmospheres, we applied the correction vectors given in
Fig.~\ref{f:ltevectors} to correct for the LTE assumption. Since for
the hottest white dwarfs these corrections become large, while 
the accuracy of temperature and gravity estimates decreases for both
LTE and NLTE analyses, one should exclude comparison of stars with
\teff$ > 70000$\,K.  This does not affect our sample, which has
a maximum temperature closer to 54000K.

Differences (after correction to NLTE) in \teff and $\log g$ between
studies for stars in common with M97, V97, and FKB are displayed in
Fig.~\ref{f:diffteff} as function of \teff.  The magnetic white
dwarfs and the binaries, which show significant contamination of the
white dwarf spectrum by the companion, are excluded.  One can now focus on
systematic differences between pairs of studies, e.g.\ our results
(NGS) versus FKB (FKB$-$NGS), V97$-$NGS, M97$-$NGS, FKB$-$M97,
M97$-$V97, and so on. However, we chose another approach and performed
an optimization that simultaneously took into account all values from
all stars in common between any samples. In other words, the values
given for the systematic differences between our study and the M97,
V97, and FKB samples form a system for direct transformation between,
say, FKB and the three other samples. The running averages computed
this way are plotted in Fig.~\ref{f:diffteff} as a solid line. The
actual value was computed in 1000\,K steps for all white dwarfs in the
respective temperature intervals \teff$\pm 10\%$.  Although this curve
does not represent the best fit to the data plotted in
Fig.~\ref{f:diffteff} alone, it is a fairly good representation of the
differences computed directly between our measurements and those of
M97, V97, and FKB. 

Since the distribution is highly non-Gaussian with many outliers, as
evident in Fig.~\ref{f:diffteff}, we decided to adopt an underlying
Lorentzian (or Cauchy) distribution for the optimization. The tails of
the Cauchy distribution are much larger than that of the corresponding
Gaussian, yielding a much lower weight for deviant points (see
discussion in Press et al.\ 1986). The dotted lines represent the
$1\sigma$ confidence interval {\em of the mean}, computed
conservatively from the rms deviations.

First, we notice a considerable scatter, larger than expected
from the internal error estimates (see the discussion below). If one
ignores the hot end, the agreement between the FKB and our temperature scale
is good; differences are below 1\%, smaller than the maximum model
differences to the Koester models (cf.\ Sect.~4), which were used by
FKB. The same atmospheres are used in
M97, and it is therefore surprising that significant
differences with M97 are present. These trends are most likely caused
by different reduction and analysis techniques. Offsets of the same order
are found in our comparison with V97, where  a different LTE
model atmosphere code is used. Although basically the same input
physics is included, this might at least partly explain those shifts in
\teff and $\log g$.

\begin{deluxetable}{lr@{$\pm$}lr@{$\pm$}lr@{$\pm$}l}
\tablewidth{0cm}
\tablecaption{SYSTEMATIC TEMPERATURE DIFFERENCES 
\label{t:dteff}}
\tablehead{
&\multicolumn{2}{c}{M97-NGS}
&\multicolumn{2}{c}{V97-NGS}
&\multicolumn{2}{c}{FKB-NGS}
\\
&\multicolumn{2}{c}{$\Delta$ \teff/\teff}	
&\multicolumn{2}{c}{$\Delta$ \teff/\teff}	
&\multicolumn{2}{c}{$\Delta$ \teff/\teff}	
}
\startdata
\teff$<30000$\,K	&$-$0.023&0.006	&0.032&0.005	&0.007&0.005\\
$30000<\teff<45000$\,K	&$-$0.013&0.003	&0.017&0.003	&0.004&0.003\\
$45000<\teff<70000$\,K	&0.036&0.006	&0.050&0.006	&0.046&0.007\\
all with $\teff<70000$\,K&$-$0.006&0.003&0.024&0.003	&0.012&0.003\\
\enddata
\tablenotetext{}{NOTES: For each temperature range, each column gives
the average difference over the respective interval, and the confidence
interval of the average difference.}
\end{deluxetable}

\begin{deluxetable}{lr@{$\pm$}lr@{$\pm$}lr@{$\pm$}l}
\tablewidth{0cm}
\tablecaption{SYSTEMATIC GRAVITY DIFFERENCES 
\label{t:dlogg}}
\tablehead{
&\multicolumn{2}{c}{M97-NGS}
&\multicolumn{2}{c}{V97-NGS}
&\multicolumn{2}{c}{FKB-NGS}
\\
&\multicolumn{2}{c}{$\Delta \log g$}
&\multicolumn{2}{c}{$\Delta \log g$}	
&\multicolumn{2}{c}{$\Delta \log g$}	
}
\startdata
$\teff<30000$\,K	&$-$0.112&0.017	&$-$0.001&0.016	&$-$0.064&0.018 \\
$30000<\teff<45000$\,K	&$-$0.086&0.017	&0.026&0.016	&$-$0.034&0.016 \\
$45000<\teff<70000$\,K	&$-$0.066&0.030	&$-$0.067&0.026	&$-$0.062&0.027 \\
all with $\teff<70000$\,K&$-$0.084&0.013&$-$0.002&0.012	&$-$0.047&0.012 \\
\enddata
\tablenotetext{}{NOTES: For details see Table~\ref{t:dteff}}
\end{deluxetable}

Our results are quantified in Tables~\ref{t:dteff}
and~\ref{t:dlogg}. We divided the white dwarfs into three groups
according to their temperature: a cool group with $\teff < 30000$\,K,
a hotter one with $30000\,\rm{K} < \teff < 45000$\,K, and the
hottest considered group with $45000\,\rm{K} < \teff <
70000$\,K. NLTE effects are still negligible in the range of effective
temperatures of the two coolest groups. Mean shifts and the confidence
range of the mean (computed as described above) are provided for these
groups and the complete sample (except stars with $\teff >
70000$\,K). The shifts discussed above and shown in
Fig~\ref{f:diffteff} are statistically significant and reach values of
$\approx$5\% in $\teff$ in the hottest bin.  In $\log g$ the
difference between our results and M97 reach $\approx$0.1\,dex for the
coolest bin. However, we emphasize that all four analyses are based on
state-of-the-art model atmospheres and $\chi^2$ fitting
techniques. Since there are no strong arguments to favor or discard
one analysis, it seems these are the systematic shifts characteristic
of modern analyses of hot white dwarfs.

\begin{deluxetable}{lrr}
\tablewidth{350pt}
\tablecaption{MEAN SCATTER $\sigma_{\rm{ind}}$ 
of TEMPERATURE and GRAVITY DETERMINATION
\label{t:scatter}}
\tablehead{
&\colhead{$\sigma(\teff)$}
&\colhead{$\sigma(\log g)$}
}
\startdata
$\teff<30000$\,K		&0.023	&0.075\\
$30000<\teff<45000$\,K		&0.023	&0.101\\
$45000<\teff<70000$\,K		&0.034	&0.128\\
all with $\teff<70000$\,K	&0.026	&0.107\\
\enddata
\end{deluxetable}

If we take the systematic shifts into account, we can use the samples
to derive estimates of the observational scatter, which can be
compared with the internal error estimates. Since we have
approximately the same scatter for different combinations of the
samples, we compute a mean scatter $\sigma_{\rm{diff}}$ for all
possible combinations (weighted by the number of stars in common). For
this purpose we correct for the systematic shifts calculated for each
of the three $\teff$ bins, whose temperature intervals are
given in Table~\ref{t:scatter}. The individual measurement errors
$\sigma_{\rm{ind}}$ add quadratically, and if we assume that the
inherent scatter is the same for all analyses (and we found no reason
to discard this assumption), the individual measurement errors can be
estimated as $\sigma_{\rm{ind}} = \frac{1}{\sqrt{2}}
\sigma_{\rm{diff}}$. Not surprisingly, the smallest scatter is
found for the ``cool'' group ($\teff < 30000$\,K) with $\sigma (\teff)
= 2.3$\% and $\sigma (\log g) = 0.07$\,dex. It increases to $\sigma
(\teff) = 3.3$\% and $\sigma (\log g) = 0.13$\,dex for the hottest
bin. This trend is expected from the statistical analysis presented in
FKB (their Fig.~1). However, the values are larger by a factor of
three or more than the internal parameter errors for a well exposed
spectrum (see e.g.\ Table~\ref{t:results}). Therefore, we conclude
that the accuracy is not limited by the noise for good spectra, and we
suggest that other effects, such as details of the extraction or
fluxing and normalization procedures, contribute more. Considering
these systematic uncertainties, the 0.3\,dex difference between the 
gravity determinations of FKB and Napiwotzki et al.\ (1993) for
HZ\,43\,A is only a $1.5\sigma$ deviation and therefore not as serious
as considered by FKB.

\subsection{Mass distribution}

\subsubsection{Derivation and Sample Comparisons}

Once the temperature and gravity of the white dwarfs are known the
mass can be determined from theoretical mass-radius relations. The
recent investigations of M97, V97, and FKB based their interpretation
on the model sequences of Wood (1995). The applied models have H- and
He-layer masses of $10^{-4} M_{\rm{WD}}$ and $10^{-2} M_{\rm{WD}}$,
respectively (``thick layers'') and a carbon core. These model
sequences are based on a set of pre-white dwarf models with end
masses of $0.6M_\odot, 0.8M_\odot$ and $0.95M_\odot$ computed by
Kawaler (see Wood 1994).  Starting models for other masses were
constructed by homology transformations. 

The mass-radius relations we use are based on the evolutionary
calculations of Bl\"ocker (1995). These models are calculated {\em ab
initio} from the main sequence. The resulting white dwarfs show
important differences compared to Wood's models. In contrast with the
carbon core models widely used for mass determination, the degenerate
core contains a mass-dependent mixture of carbon and oxygen (and trace
elements). However, note that Wood (1995) also presented some white
dwarf sequences with C/O cores. The H- and He-layer masses resulting
from evolutionary calculations depend on the stellar mass (Bl\"ocker
1995; Bl\"ocker et al.\ 1997). While the canonical thick hydrogen
layer mass of $10^{-4} M_{\odot}$ is met for a $0.6 M_{\odot}$ white
dwarf, it decreases from several $10^{-3} M_{\odot}$ for $0.3
M_{\odot}$ down to $10^{-6} M_{\odot}$ for $1.0 M_{\odot}$. This
results in a mass-radius relation which is steeper than a relation
based on a constant layer mass. Consequently, the derived mass
distribution will be narrower, if evolutionary layer masses are used.

The lower mass limit for a C/O core white dwarf is $0.46 M_{\odot}$,
the limiting mass for central helium burning 
in low mass stars (Sweigart et al.\ 1990). 
Thus, white dwarfs with a
lower mass possess a helium core. At the current age of the universe
He white dwarfs have not been produced by single star evolution,
but are the result of binary evolution where the
hydrogen-rich envelope was stripped away along the (first) red giant
branch (Kippenhahn 1967; Iben \& Tutukov 1986).  We use the recent
evolutionary models of He white dwarfs calculated by Driebe et al.\ (1998)
based on this scenario. 

The combined tracks of Bl\"ocker (1995) and Driebe et al.\ (1998) cover the
mass range $0.18M_\odot$ to $0.94M_\odot$. We supplemented this set
with the 1.0, 1.1, and $1.2M_\odot$ carbon core sequences of Wood
(1995) with ``thin'' layers. Since the hydrogen envelope mass decreases
with increasing white dwarf mass in Bl\"ocker's models, their highest mass
models effectively correspond to ``thin layer'' models.
Since high mass white dwarfs are already close to the
zero-temperature configuration, the remaining departures from
consistent evolutionary calculations can be ignored for practical
purposes.

The position of the analyzed white dwarfs in the temperature/gravity
plane is shown in Fig.~\ref{f:tracks} along with the tracks used for
the mass determination. The individual white dwarf masses are given in
Table~\ref{t:results}, and the resulting mass distribution is shown in
Fig.~\ref{f:mass}. We have also redetermined masses for the BSL sample
with the Bl\"ocker/Driebe mass-radius relations and supplement 
Fig.~\ref{f:mass} with the resulting distribution.

Our mass distribution possesses the same basic features as found in
the EUV selected samples of M97, V97, and FKB. The sharp peak centered
at $\approx 0.59M_\odot$ is in principal agreement with the earlier
investigations of KSW, Weidemann \& Koester (1984), and BSL, with a
sharp falloff towards lower masses, and a less steep decline towards
higher masses with a tail of high mass white dwarfs. Considering the
relatively large observational scatter we found in Sect.~\ref{s:comparison}, 
the {\em underlying} distribution may be extremely sharp. The mean mass is
$0.67 M_\odot$. However, this value is strongly biased by the
few very high mass white dwarfs. 

We decided to follow the recipe of FKB and fitted the mass peak with a 
Gaussian. Although a Gaussian is not a good representation of the white dwarf 
mass distribution, it gives a robust estimate of the peak mass. The fit result 
is, in principle, dependent on the adopted binning of the mass intervals. 
However, in some test calculations we found effects exceeding a few 
$0.001 M_\odot$ only if the bin width was larger than $0.05 M_\odot$. The 
values given in this paper were derived by fitting Gaussians to virtually 
unbinned mass distributions (formal bin width $0.001 M_\odot$). 

We were concerned by the possibility that the different fraction of low and 
high mass white dwarfs found in different samples (see discussion 
below) may skew the Gaussian fit of the main peak to higher or lower masses. 
We tested this by fitting the BSL, and FKB and our mass distributions with 
multi-component Gaussians, which fitted the secondary high and low mass peaks 
separately. We found no deviation higher than $0.003 M_\odot$ in any case and 
concluded that the single Gaussian fit of the main peak yields a rather stable 
estimate.

With this
method, we derived a peak mass of our sample of $0.589 M_\odot$. We reanalysed 
the FKB, M97, and V97 samples with our mass-radius relations. We applied the 
corrections to NLTE and excluded the white dwarfs with temperatures in
excess of  
70000\,K. The peaks masses are $0.555 M_\odot$ (FKB), $0.535 M_\odot$ (M97),
and 0.582 (V97). These differences reflect the systematic differences in 
$\teff$ and $\log g$ determinations discussed above and probably different 
selection criteria in the case of FKB.

The probable He core white dwarf RE\,J0512-004 is the only object in our sample
with a mass below $0.5M_\odot$, but four white dwarfs have 
masses in excess of $1.0M_\odot$.
The frequency of low and high mass white dwarfs found in our and other EUV
selected samples is qualitatively
different from that found in optically selected samples such as BSL. 
BSL detected in their sample of 129 white dwarfs 16 objects with masses  
below the limiting mass for C/O core white dwarfs and a total of 28 white 
dwarfs with $M<0.5M_\odot$. These numbers are reduced to 10 and 16 
respectively, if we redetermine the masses with the mass-radius
relations of Bl\"ocker (1995) and Driebe et al.\ (1998). However,
that is still a much higher fraction than we find in our EUV selected
sample. 

\subsubsection{Cooling Rates and Detection Probability}

Why do optical and EUV selected samples contain different 
fractions of high and low mass white dwarfs?
All helium white dwarfs from the BSL sample have temperatures below 
30000\,K; the only helium white dwarf candidate in our sample has $\teff 
\approx$32000\,K. This implies that the detection probability of
low mass white dwarfs in EUV selected samples is much lower, because the
fraction of low mass white dwarfs at higher $\teff$ is much lower 
than it is for $\teff<30000$\,K. This viewpoint is supported by 
the helium white dwarf sequences calculated by Driebe et al.\ (1998). 
We have plotted the variation of temperatures with time for 
white dwarfs with helium and C/O cores in Fig.~\ref{f:tevol}. 
The detection probability in a given evolutionary stage goes as the
inverse of the rate of temperature decrease at that stage. New He
white dwarfs cool down rapidly to a certain (mass dependent)
temperature, whereafter the cooling rate drops drastically and the star
remains on a temperature  plateau for a long time. This behavior is
most pronounced for the lowest masses. Hydrogen shell burning plays an
important role 
for this plateau phase. Although the energy production by  
hydrogen burning drops down dramatically after the star enters the
white dwarf sequence, it still produces a significant fraction of the 
total luminosity, enough to bring the cooling to a standstill.
We can conclude from Fig.~\ref{f:tevol} that it is extremely unlikely to 
detect a helium white dwarf during its first rapid cooling phase. It is
easier to find them during their plateau phase or afterwards. This means
there is a strong bias towards detecting helium white dwarfs in 
samples of relatively ``cool'' DA stars (like BSL) instead of the 
overall much hotter EUV selected samples.

We can explain in a similar fashion why high mass white dwarfs are found
preferentially in the $30000\,\rm{K} < \teff < 50000$\,K range. Due to the
drop in neutrino cooling, the evolutionary speeds of massive white dwarfs
are relatively low in this range, as demonstrated by the $0.940M_\odot$
track, the most massive stellar remnant calculated by Bl\"ocker (1995). 
This increases the detection probability of more massive white dwarfs
in EUV selected samples.  

\subsubsection{Interstellar Absorption}

Even if we restrict our comparisons to hot subsamples of white dwarfs, the
fraction of massive white dwarfs is much higher in the EUV selected samples
than it is in optically selected ones. An extensive discussion of this problem 
is carried out by FKB. FKB argue that the number of high mass white dwarfs
discovered per sky area is similar for the whole sky EUV surveys and the
optical Palomar-Green (PG) survey (Green et al. 1986), which covered
roughly 25\%  
of the sky. In other words, a whole sky version of the PG survey should have
discovered all but one of the EUV detected white dwarfs with $M>1.1M_\odot$. 
At a given temperature, white dwarfs with lower masses are larger,
consequently more luminous, and detectable to larger distances.
For instance at 30000K, a white dwarf of $0.5M_\odot$ is detectable at
distances $\sim 30\%$ larger than a $0.7M_\odot$ white dwarf.  At the
maximum distance sampled by, e.g., the PG survey, the probability is
very high that interstellar matter effectively absorbs all EUV 
radiation.  Thus the dominant effect in EUV surveys is not a selection
for massive white dwarfs as suggested by V97, but a selection {\em
against} low mass white dwarfs, due to a sample volume strongly
affected by interstellar absorption.  

\subsubsection{Temperature Dependence of the Derived Mass Peak}

FKB noted a moderate apparent trend of the white dwarf peak
mass with temperature. Since their sample also included optically selected
white dwarfs, they analyzed a considerably higher fraction of stars
with $\teff < 25000$\,K and therefore have a larger temperature
baseline than pure EUV samples.  Analogous to this finding, V97 found
an $0.03M_\odot$ offset between the peak of the mass distribution in
their EUV selected sample and the (cooler) white dwarfs analyzed by
BSL. In both cases this temperature dependence decreases if Wood's
models with ``very thin'' layers ($\mhe = 10^{-4}M_{\rm{WD}}$, no
hydrogen layer) instead of the canonical thick layer models 
($\mhe = 10^{-2}M_{\rm{WD}}, \mh = 10^{-4}M_{\rm{WD}}$)
are used. However, part of the trend noted in V97 may be related to
the systematic 
differences between the analyses discussed in Sect~5.3.  Since V97
compared results from two different analyses, it is not possible to
judge if the offset is real or just an artifact. The case of FKB is
stronger, because they based their case on a homogeneously analyzed
sample. FKB warned that the temperature dependence of the mass
distribution peak may be caused by inadequacies remaining in the model
atmospheres. A more detailed discussion is given in the next section.

Some evidence that part of the temperature trend stems from the
analyses comes from our intercomparisons presented in Sect~5.3. We
showed that systematic differences between the four investigated
samples exist that vary with effective temperature. This might mimic
a temperature dependence of the sample peak mass. A gravity offset of
0.1\,dex transforms into mass offsets of 0.050$M_\odot$,
$0.044M_\odot$, and $0.034M_\odot$ for a $0.6M_\odot$ white dwarf with
25000\,K, 40000\,K, and 60000\,K, respectively. Even a relatively
small systematic $\log g$ difference of 0.05\,dex (cf.\
Table~\ref{t:dlogg}) corresponds to a $0.02M_\odot$ offset.  The
scatter $\sigma$ of {\em individual} gravity determinations reported
in Table~\ref{t:scatter} corresponds to $\sigma(M) \approx
0.04M_\odot$, nearly independent of $\teff$.

\subsubsection{Cooling Tracks}

Another part of the $\teff$ dependence of the
mass peak may be caused by the use of Wood's cooling tracks by FKB and V97.
These tracks are not completely self-consistent since not all model sequences
were not calculated {\em ab initio} from the main sequence. This is not 
important for cooler white dwarfs but may cause deviations from {\em ab initio}
tracks especially for the hottest white dwarfs, where structure depends
sensitively on the evolutionary history (Bl\"ocker \& Sch\"onberner 1990). 

We redetermined the masses of the BSL white dwarfs with the Bl\"ocker/Driebe
mass-radius relations and derived a peak mass of $0.559M_\odot$ . 
That is $0.03M_\odot$ lower than our peak mass, but having in mind the range
of mass determination derived for the EUVE selected samples, we cannot 
consider this a significant difference. FKB divided their sample into cool
($\teff<35000$\,K) and a hot (35000\,K $< \teff < $75000\,K) subsamples and
derived a $0.029M_\odot$ higher peak mass for the hot sample. Our reanalysis
of the FKB LTE results with the Bl\"ocker/Driebe mass-radius relation yields
virtually the same offset: $0.025M_\odot$. However, the difference is brought
down to $0.010M_\odot$ ($0.550 M_\odot$ vs.\ $0.560 M_\odot$) if we apply 
corrections to NLTE. One can imagine that a difference of this order
(if significant at all) can easily be produced by our  neglect of
metallicity effects (cf.\ Lanz et al.\ 1996, Barstow et al.\ 1998).

Therefore, given the combination of sample selection
and systematic effects in analyses to date,
our results do not confirm the presence of intrinsic, systematic
mass differences between hot and cool white dwarfs.

\section{Conclusions}

We have obtained temperatures, gravities, and masses for a sample of 46 
extreme ultraviolet selected DA white dwarfs. These data complement a 
near infrared survey for low mass companions. The stellar parameters were 
determined by fitting the hydrogen Balmer line profiles with NLTE model 
spectra. A map of LTE correction vectors was constructed, which allows the 
transformation of LTE results to the NLTE scale. These shifts become important 
for temperatures above 50000\,K (see also Napiwotzki 1997). 

Three recent analyses of DA white dwarfs have a considerable overlap
with our sample and allow a direct check for systematic errors and
individual measurement scatter. Although all four analyses applied
similar model atmospheres and fitting techniques, we recognized
systematic shifts up to 5\% in temperature and 0.1\,dex in gravity
determinations. These systematic errors have to be considered when the
results are interpreted and compared to other studies. The individual
measurement errors increase with temperature: $\sigma (\teff)$ from
2.3\% to 3.3\% and $\sigma (\log g)$ from 0.07\,dex to 0.13\,dex. This
is larger than the typical errors computed with $\chi^2$ fit
procedures for a well-exposed spectrum and indicates that accuracy is
limited by noise in combination with other effects such as details of
the extraction, fluxing or normalization procedures. It is unclear if
these $\sigma$ values indicate lower limits, or scale (at least
partially) with the S/N ratio of the spectra. It will be a challenge
to resolve this question. Repeated observations performed by one
observer, at one telescope, reduced the same way and analysed with one
method as performed by BSL, FKB, and us (Table~2) do not reveal all
effects. Independent observations analysed independently are
necessary.

Masses have been inferred from theoretical mass-radius relations based on
the evolutionary calculations of Bl\"ocker (1995) for C/O white dwarfs and
Driebe et al.\ (1998) for He white dwarfs. An important feature of these
calculations are the hydrogen and helium layer masses which depend on stellar
mass. We find a sharp peak centered at $\approx$$0.59M_\odot$ 
in agreement with the previous investigations of
KSW and BSL.  We redetermined masses of
the BSL white dwarfs with the Bl\"ocker/Driebe mass-radius relation and derived
a peak mass of $0.56M_\odot$. At first glance this seems to confirm 
the systematic offset reported by V97. However, this offset could be
explained by 
systematic differences of the model atmosphere analyses as well. 
Our reanalysis of the homogeneous FKB sample showed that the temperature 
dependence of the mass peak nearly vanishes, when corrected for their
LTE assumption.  We conclude that the observational data
presented here and in similar studies are well explained by canonical
stellar evolution theory, i.e. white dwarfs with thick envelopes. 

We find only one object (RE\,J0512-004) with a mass below $0.5M_\odot$, 
which is a possible helium core white  dwarf, but four white dwarfs with
masses in excess of $1.0M_\odot$. This ratio of high-
and low-mass white dwarfs is quite different from that found by BSL. 
This can partly be understood as the result of variable evolutionary 
time scales of high and low mass white dwarfs. We agree with FKB that
another part can be explained by a selection {\em against} low mass
white dwarfs in EUV selected samples. 

The principal aim of our project is the search for and study of cool
main sequence companions of our white dwarf sample. Five white dwarfs
already show red contamination in our optical spectra. A comprehensive
discussion of these  and other binaries will be given in a forthcoming
paper presenting IR photometry for the white dwarf sample.  In this
article we provided the basic white dwarf data necessary to interpret
our binary sample.   

\acknowledgments We thank T.\ Driebe, F.\ Herwig, and T. Bl\"ocker for
providing us with their tracks and computing the white dwarf masses and D.\
Koester, who made some model spectra available for our model comparison. We are
grateful to Steward Observatory for an unexpected, generous award of six nights
of unclaimed 2.3-m time, which made possible simultaneous optical and infrared
observations of our sample of stars. We thank Perry Berlind for obtaining
several white dwarf spectra. PJG acknowledges support through NASA Contract
NAS8-39073 (ASC).

\clearpage

\figcaption[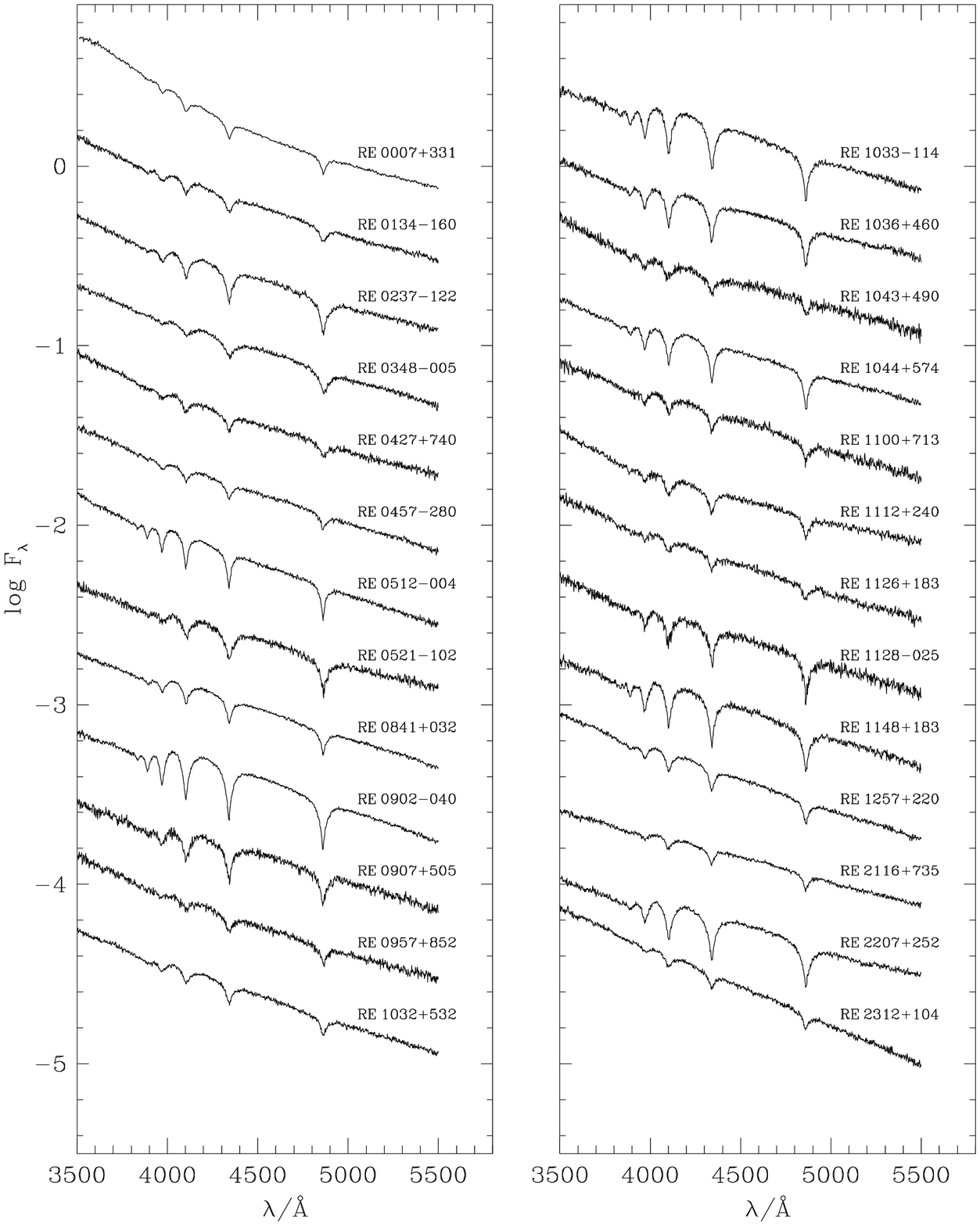]{Spectra of the white dwarfs observed at Steward 
	observatory \label{f:steward}}

\figcaption[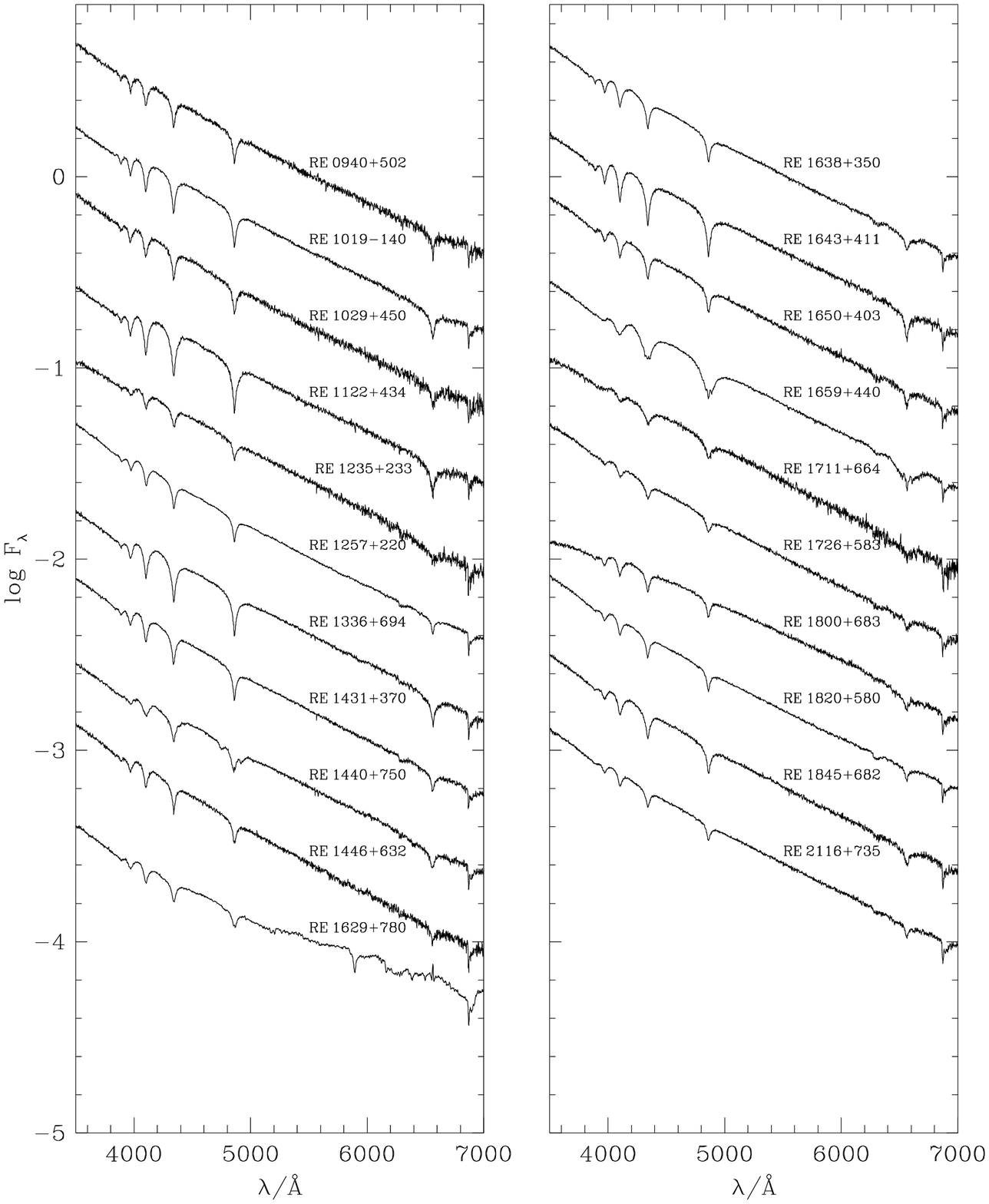]{Spectra of the MMT sample \label{f:MMT}}

\figcaption[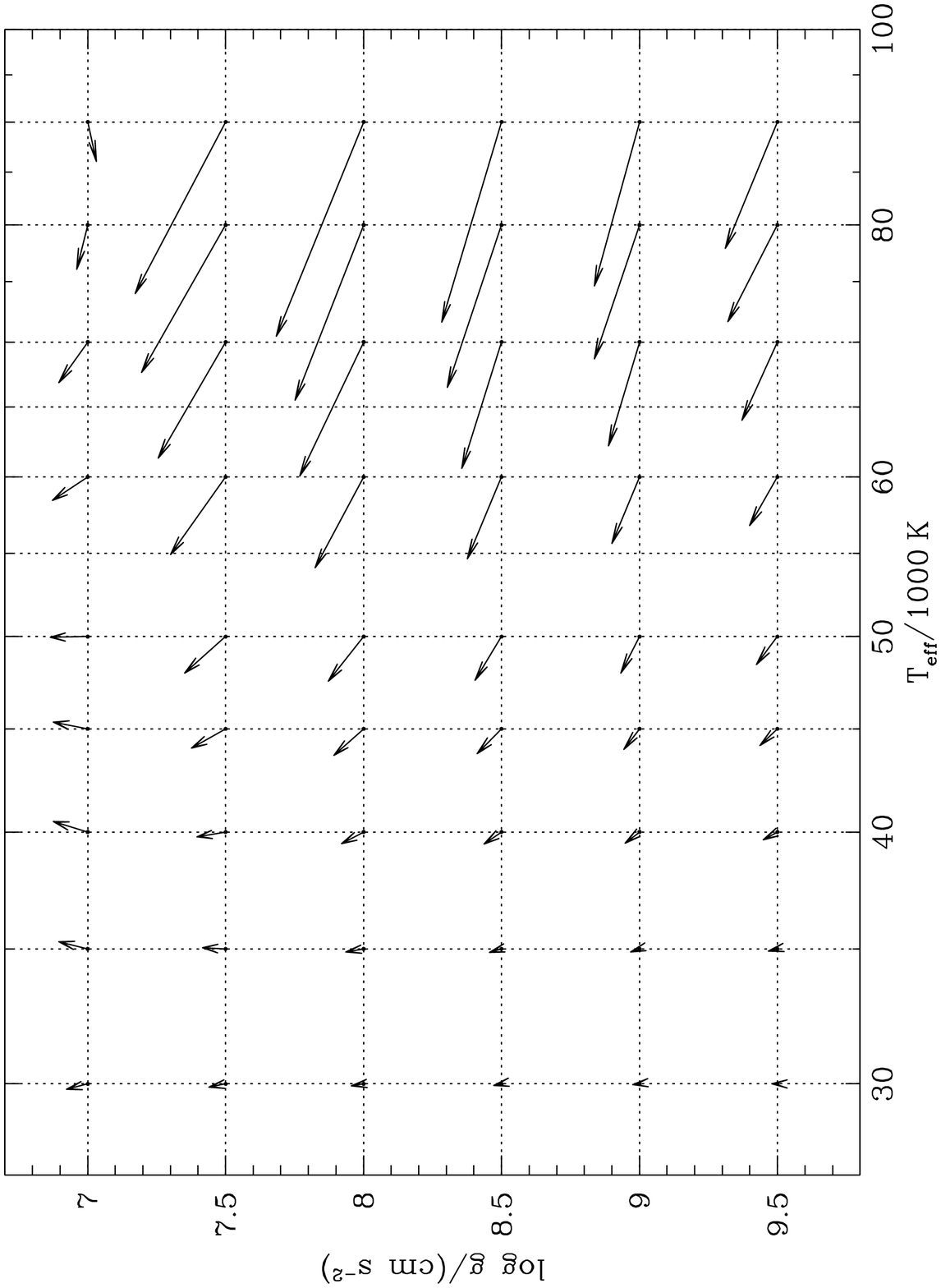]{LTE offsets. The differences are magnified
	three times.  	The vectors give the correction which must be
	applied to transform LTE results to the NLTE scale.
	\label{f:ltevectors}}

\figcaption[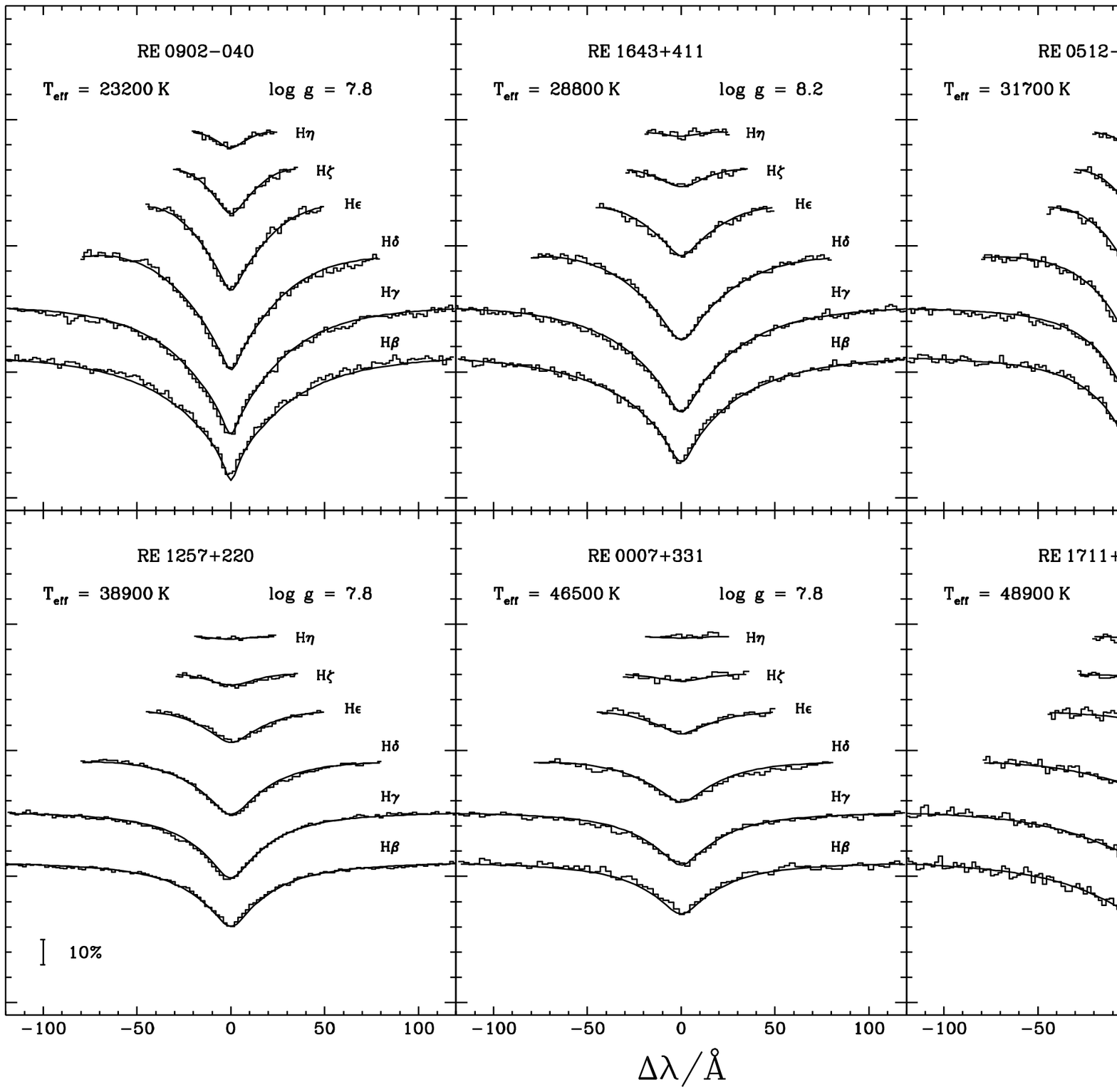]{Balmer line fits for a representative set of 
	white dwarfs. \label{f:sampleres}}

\figcaption[compspek.eps]{Flux calibrated spectrum of the DA+M pairs 
	RE\,J0134-160, RE\,1036+460, RE\,J1043+490, RE\,J1126+183, 
	and RE\,J1629+780. From top to bottom we show the composite spectrum, 
	the best fit of the white dwarf, and the resulting spectrum of the M 
	star. The spectrum of the M star companions of RE\,J0134-160 and
	RE\,J1126+183 are multiplied by 3.
	\label{f:compspek}}

\figcaption[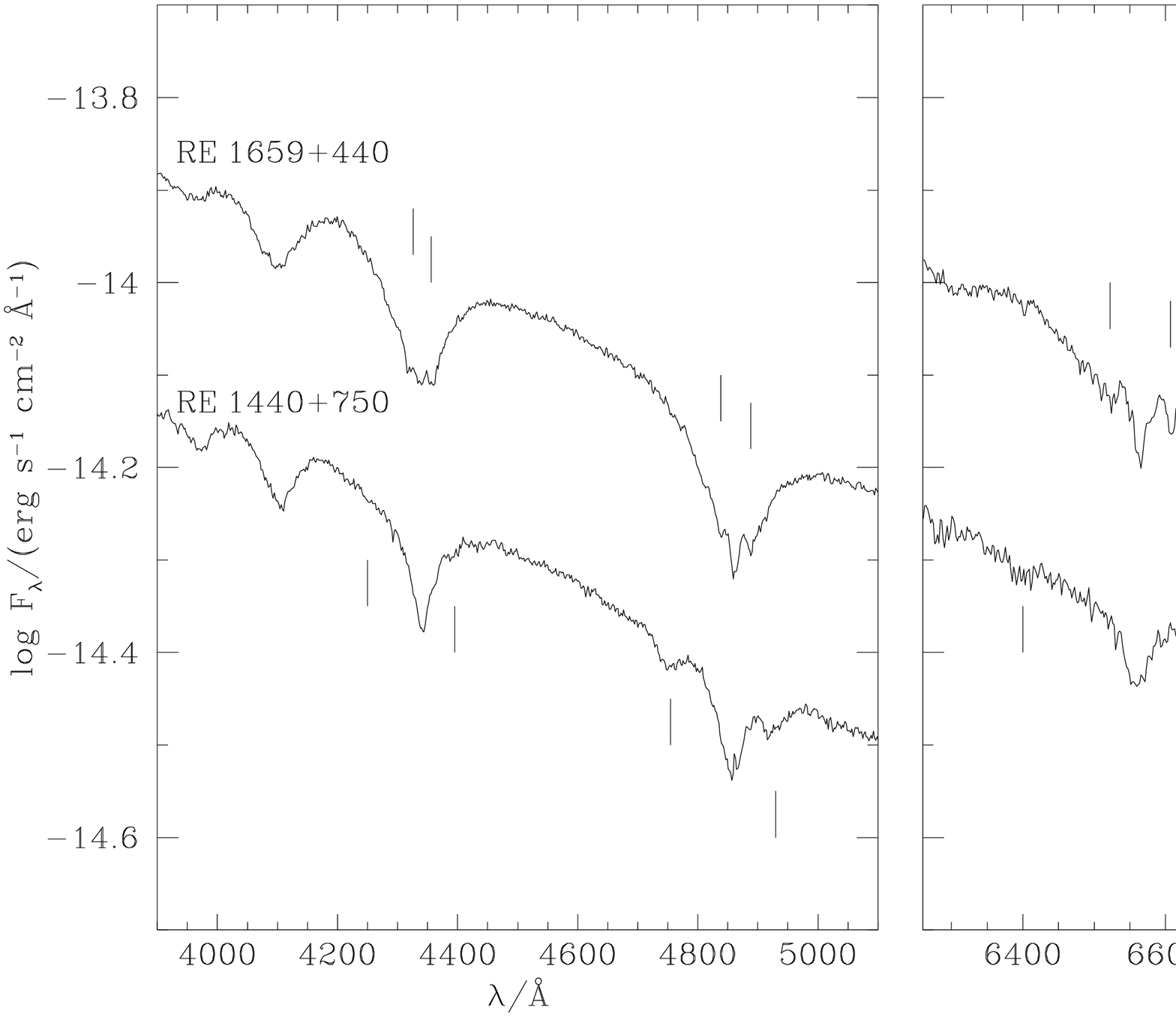]{Spectra of the magnetic white dwarfs PG\,1658+441 and
	RE\,J1440+750. The position of the $\sigma$ components of the Balmer
	lines are indicated. 
	\label{f:magwd}}

\figcaption[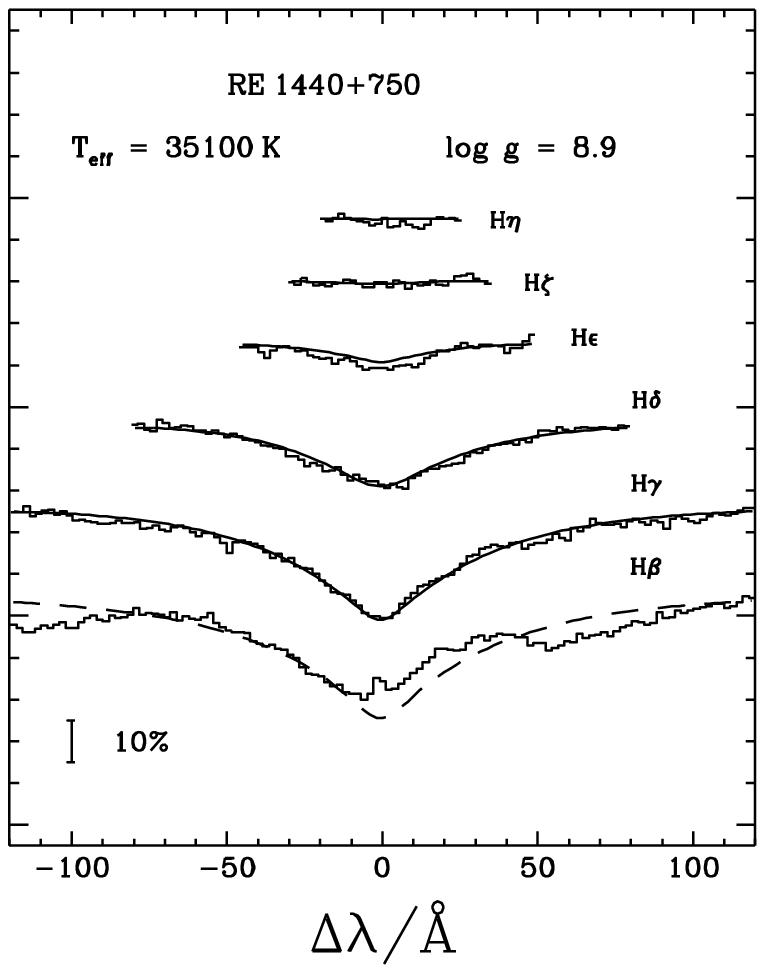]{Best fit of the magnetic white dwarf  RE\,J1440+750.
	As indicated by the dashed line, \hbeta\ was not used for the 
	parameter determination.
	\label{f:re1440}}

\figcaption[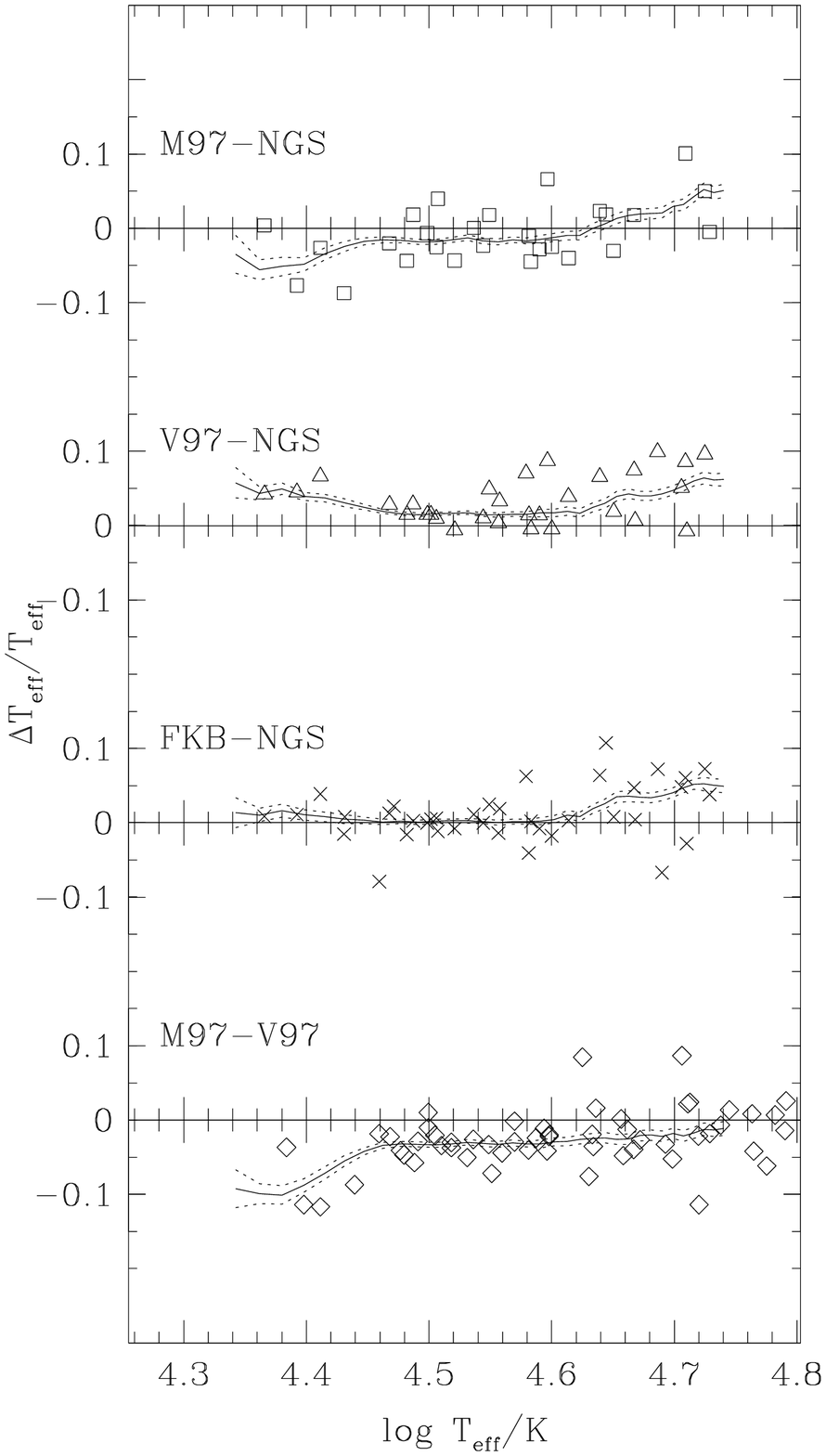]{Differences in temperature (left), and 
	gravity (right) between the M97, V97, and FKB samples and our
	own on a star by star basis. The same comparison is carried
	out for the LTE studies of M97 and V97 in the bottom panel.
	The smoothed average of the differences is plotted with solid
	lines, while the standard error of the mean difference is
	indicated by the dotted lines. The $\teff$ values used for the
	x-axis are NLTE from this paper (NGS), except for the bottom
	panel, which uses V97 (corrected to NLTE).
	\label{f:diffteff}}

\figcaption[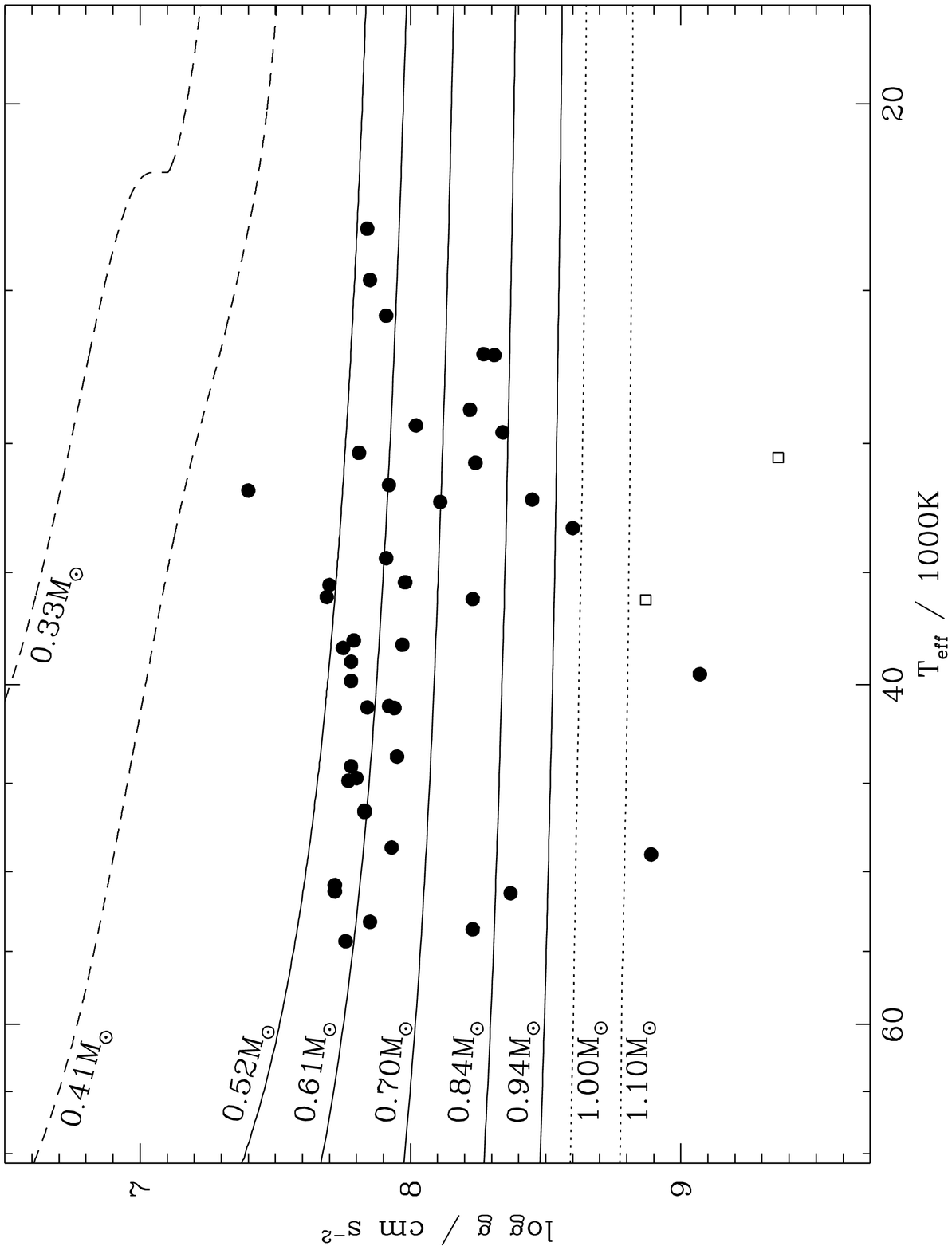]{Effective temperature and gravity of our white dwarf
	sample compared with evolutionary tracks. Solid lines are
	Bl\"ocker (1995); dashed lines are Driebe et al. (1998);
	dotted lines Wood (1995). The magnetic white dwarfs are marked
	by open symbols.
	\label{f:tracks}}

\figcaption[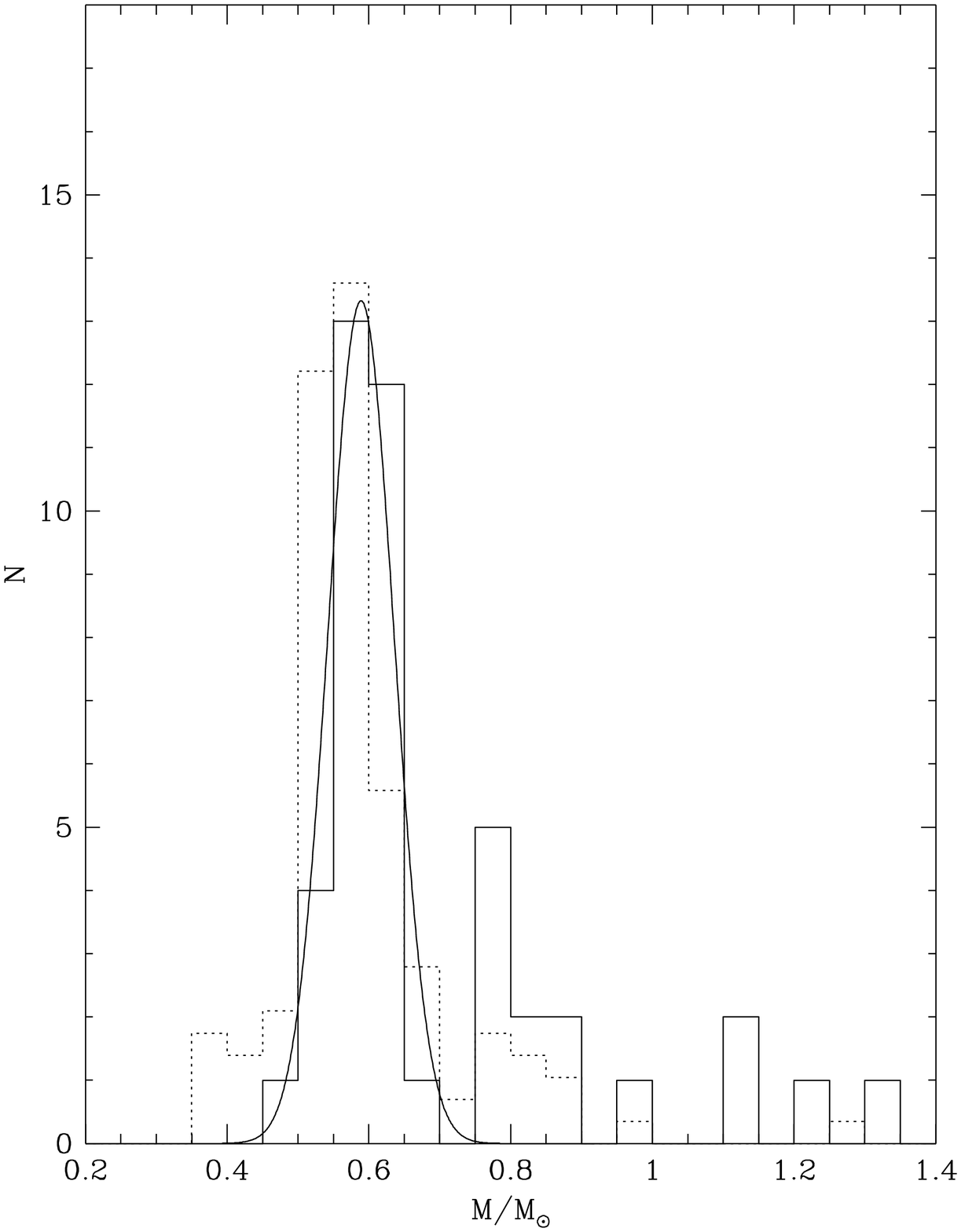]{Mass distribution of white dwarfs (binned over 
	$0.05\,M_{\odot}$). The solid lined histogram indicates the result 
	for our sample. The dotted line is the BSL sample with masses 
	redetermined from the Bl\"ocker/Driebe relations (normalized to the same
	sample size). The Gaussian curve represents the best fit to our mass 
	distribution as described in the text.
	\label{f:mass}}

\figcaption[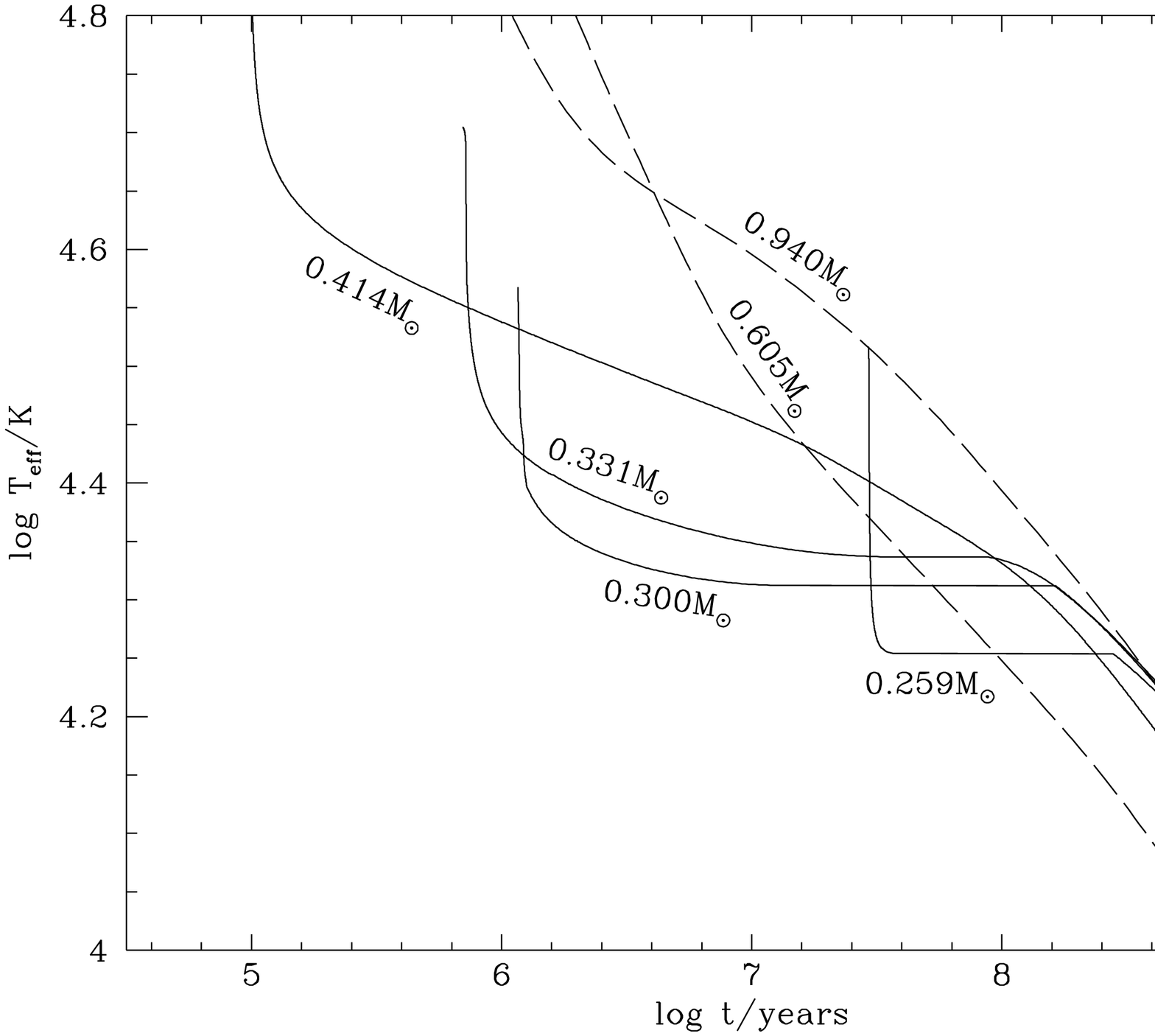]{Time evolution of stellar surface temperature for 
	helium core white dwarfs (tracks from Driebe et al.\ 1998; solid lines)
	and C/O white dwarfs (tracks from Bl\"ocker 1995; dashed lines).
	The tracks are labeled with the stellar masses.
	\label{f:tevol}}


\clearpage

\begin{figure}
\epsscale{0.70}
\plotone{steward.eps}
\end{figure}
\clearpage

\begin{figure}
\epsscale{0.70}
\plotone{mmt.eps}
\end{figure}

\clearpage

\begin{figure}
\epsscale{0.70}
\plotone{lteshift.eps}
\end{figure}

\clearpage

\begin{figure}
\epsscale{0.80}
\plotone{sample.eps}
\end{figure}

\clearpage

\begin{figure}
\epsscale{0.80}
\plotone{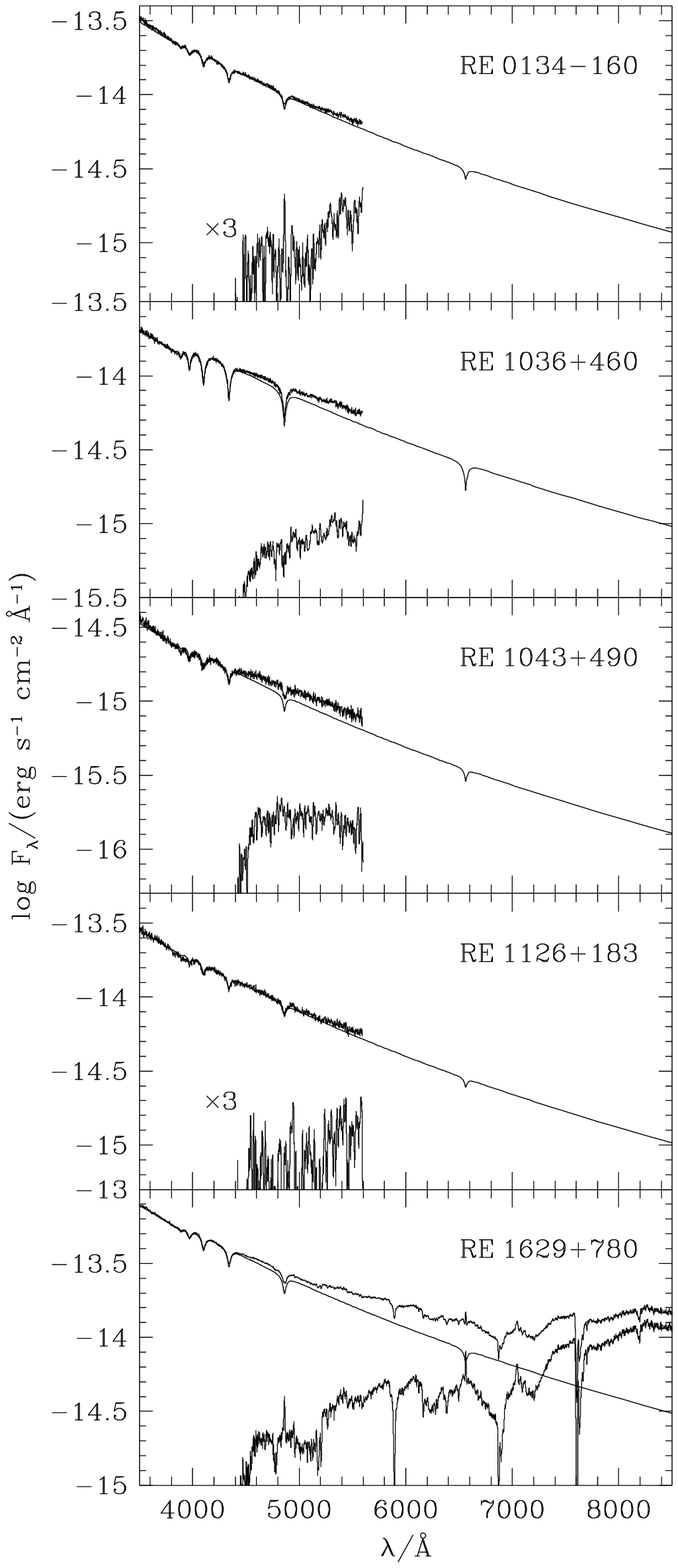}
\end{figure}
\clearpage

\begin{figure}
\epsscale{0.80}
\plotone{magwd.eps}
\end{figure}

\clearpage

\begin{figure}
\epsscale{1.00}
\plotone{re1440.eps}
\end{figure}

\clearpage

\begin{figure}
\epsscale{0.70}
\plottwo{dteff.eps}{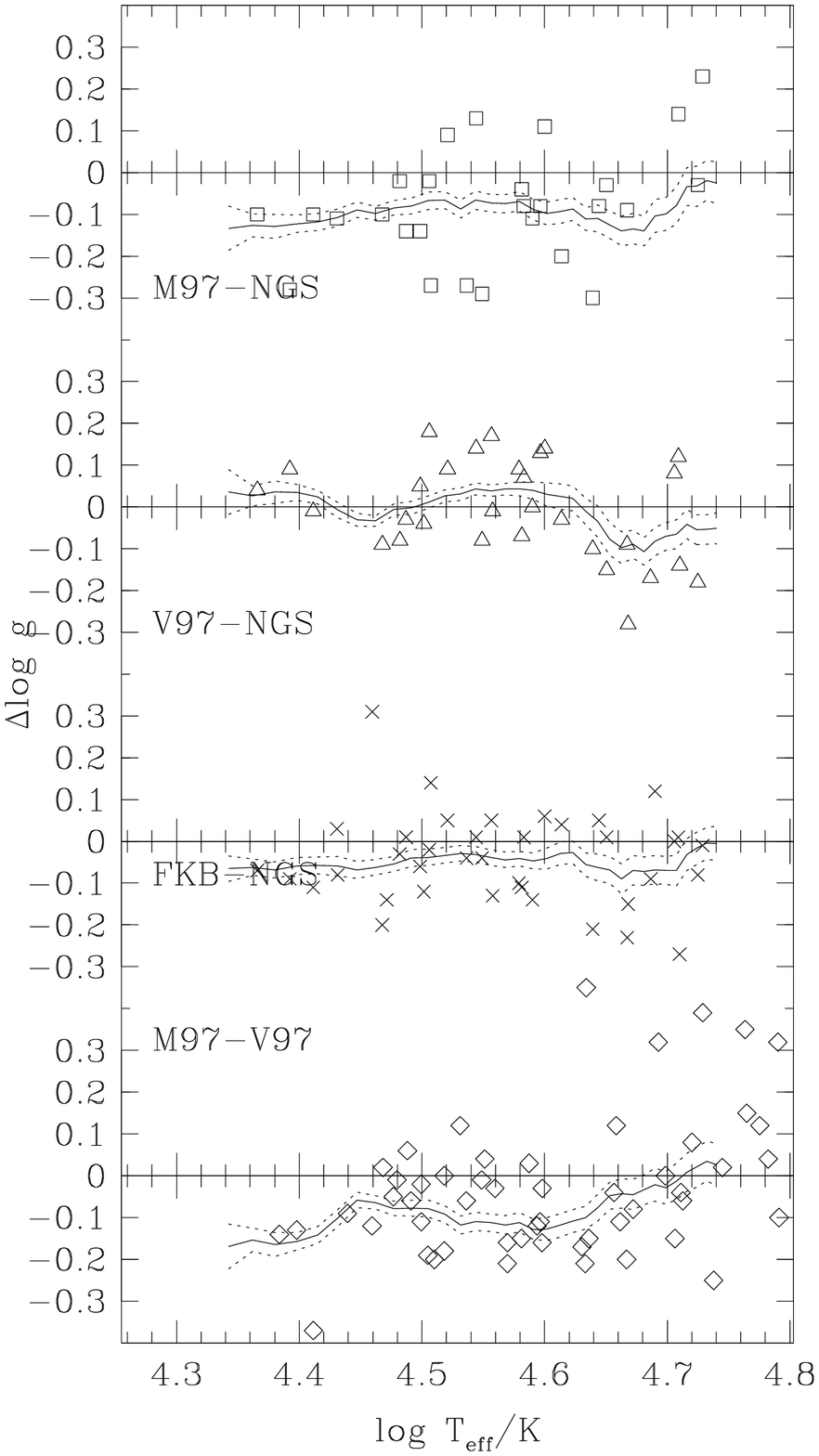}
\end{figure}

\clearpage

\begin{figure}
\epsscale{0.70}
\plotone{tracks.eps}
\end{figure}

\clearpage

\begin{figure}
\epsscale{0.70}
\plotone{histo.eps}
\end{figure}

\clearpage

\begin{figure}
\epsscale{0.70}
\plotone{tevol.eps}
\end{figure}

\end{document}